\documentclass[referee,sn-basic]{sn-jnl}

\usepackage{graphicx}%
\usepackage{multirow}%
\usepackage{amsmath,amssymb,amsfonts}%
\usepackage{amsthm}%
\usepackage{mathrsfs}%
\usepackage[title]{appendix}%
\usepackage{xcolor}%
\usepackage{textcomp}%
\usepackage{manyfoot}%
\usepackage{booktabs}%
\usepackage{algorithm}%
\usepackage{algorithmicx}%
\usepackage{algpseudocode}%
\usepackage{listings}%
\usepackage{ulem}%

\theoremstyle{thmstyleone}%
\theoremstyle{thmstyletwo}%

\theoremstyle{thmstylethree}%

\begin{document}

\title[Tracking the normal modes of a bridge using DAS]{Tracking the normal modes of an overpass highway bridge using Distributed Acoustic Sensing}


\author*[1]{\fnm{E. Diego} \sur{Mercerat}}\email{diego.mercerat@cerema.fr}
\equalcont{These authors contributed equally to this work.}

\author[1]{\fnm{Martijn} \spfx{van den} \sur{Ende}}
\equalcont{These authors contributed equally to this work.}

\author[1]{\fnm{Anthony} \sur{Sladen}}
\equalcont{These authors contributed equally to this work.}

\author[1]{Vanessa Carrillo Barra}
\author[2]{Julian Pel\'aez Quiñones}
\author[1]{Daniel Mata Flores}
\author[1]{Philippe Langlaude}
\author[3]{Destin Nziengui}
\author[4]{Olivier Coutant}

\affil[1]{Université Côte d'Azur, CEREMA, CNRS, IRD, Observatoire de la Côte d'Azur, Géoazur, Valbonne, France}
\affil[2]{Dept. of Physics and Technology. University of Bergen, Norway}
\affil[3]{Febus Optics, Pau, France}
\affil[4]{ISTerre, Université Grenoble Alpes - CNRS, Grenoble, France}


\abstract{Distributed Acoustic Sensing (DAS) of ambient vibrations is a promising technique in the context of structural health monitoring of civil engineering structures. The methodology uses Rayleigh backscattered light from small deformations at different locations of the sensed fiber-optic cable, turning it into a large array of equally distributed strain sensors. In this paper, we demonstrate the feasibility of using DAS technology to record dynamic strain used for modal identification through the Operational Modal Analysis (OMA) of a strut-frame bridge overpassing the A8 highway in southeastern France. Modal identification using DAS data is successful despite its predominantly axial sensitivity (along fiber), though the help of three-component seismometers is useful for discriminating the main motion direction of each identified mode. The identification of bridge's normal modes with unprecedented spatial resolution is obtained from the lowest (transverse and longitudinal) modes to high-order modes that present significant vertical motion. In addition, strong seasonal effects are observed in both the absolute frequency values and the modal shapes of the first transverse and longitudinal modes of the bridge, comparing ambient vibration testing and DAS surveys carried out in the summer and winter periods.}




\maketitle

\section{Introduction}\label{intro}

Operational Modal Analysis (OMA) is a non-invasive technique that has been commonly used for monitoring the structural health of bridges and civil engineering structures since the end of the 1990s \citep{peetersOneyearMonitoringZ24Bridge2001a,peetersVibrationbasedDamageDetection2001}. It has been used both for the dynamic characterization of structures (in terms of modal frequencies, modal shapes, and damping coefficients) and for structural health monitoring using dynamic characteristics and particularly their variations over time, to detect and localize structural damage \citep{sawickiLongtermStrainMeasurements2020,gaoOptimalLayoutSensors2020}. Previous works\citep{abdelwahabDAMAGEDETECTIONBRIDGES1999,reyndersDamageIdentificationTilff2007,ghahremaniBridgeHealthMonitoring2022} have proposed the use of strain measurements to estimate modal shape curvatures for damage identification. The accuracy in calculating the curvature is substantially improved by directly measuring dynamic strain rather than deriving the modal curvatures from velocity or acceleration measurements\citep{reyndersDamageIdentificationTilff2007,anastasopoulosInfluenceDamageTemperature2019}. However, those previous studies involved only strain gauge measurements at relatively few locations on the bridge deck. To precisely constrain the model characteristics and to potentially detect damage formation, it is necessary to deploy a large number of strain sensors distributed over the structure under investigation. 

As an alternative to conventional seismometers and strain gauges, Distributed Acoustic Sensing (DAS) technology converts a standard optical fiber into a dense array of strain(-rate) sensors. This revolutionary metrological approach, which nowadays is widely used in applied seismology \citep{joussetDynamicStrainDetermination2018,vandenendeEvaluatingSeismicBeamforming2021,liorImagingUnderwaterBasin2022,trabattoniMicroseismicityMonitoringSite2022} and earthquake engineering \citep{gorshkovScientificApplicationsDistributed2022,kishidaDistributedOpticalFiber2022,liuTurningTelecommunicationFiberOptic2023}, allows measurements with unprecedented spatial density and at sampling frequencies compatible with the monitoring of dynamic deformation of engineering structures (from tens to hundreds of Hertz). The dense and distributed nature of DAS is highly desirable for OMA applications that require high spatial resolution, such as for modal shape characterization. 

Within the framework of Structural Health Monitoring (SHM), some recent works have turned to analyze strain data acquired by DAS and other fiber-optic based techniques from bridges and other engineering structures, although the analysis is still focused on estimating resonance frequencies \citep{liDetectabilityBridgeStructuralDamage2020,hubbardDynamicStructuralHealth2021,reyndersVibrationMonitoringRailway2021,abedinStructuralHealthMonitoring2023} but the complete modal shape identification is still in early stages : some previous research used relatively limited number of channels \citep{lienhartDistributedVibrationMonitoring2023}, concentrated exclusively on the fundamental bending modes \citep{monsbergerDistributedFiberOptic2021,strasserStaticDynamicBridge2023}, while others have restricted themselves to linear array geometries \citep{liuTurningTelecommunicationFiberOptic2023,petladwalaStatisticalStudyBridge2023,rodetUrbanDarkFiber2024}.

To further explore the feasibility of DAS for OMA of medium- and long-span bridges, the main objective of this paper is to characterize the modal shapes and frequencies of an overpass highway bridge located near the city of Nice in the southeast of France. The advantage of DAS over conventional instrumentation is characterized by the high spatial resolution of the experimental modal shapes obtained. One novelty of the present work lies in the identification of eight normal modes of the bridge under study thanks to the use of two parallel optic fibers with submetric spatial sampling. Furthermore, by comparison of the DAS-derived results from two different surveys carried out in winter and summer periods, we observe a clear seasonal effect that drastically alters the bridge dynamics. In fact, the first longitudinal mode almost completely disappears during summer time due to thermal expansion of the deck that impedes its longitudinal motion. Although long-term continuous structural monitoring with point-wise instrumentation is challenging, these results show that DAS offers a feasible alternative, allowing operators to perform continuous safety assessments. 

In the following, we present the Methodology Section that encompasses first the description of both instrumentation surveys, and secondly the data analysis proposed. After that, the OMA results from both the DAS and the classical seismometer datasets are presented and discussed. We end up with a discussion about the seasonal effect on boundary conditions at the bridge abutments that clearly impact the frequency values and the modal shapes estimated in winter and summer periods.

\section*{Methodology}

\subsection*{Instrumentation surveys}

\begin{figure}[t]
\centering    
(a)\includegraphics[width = 0.9\textwidth]{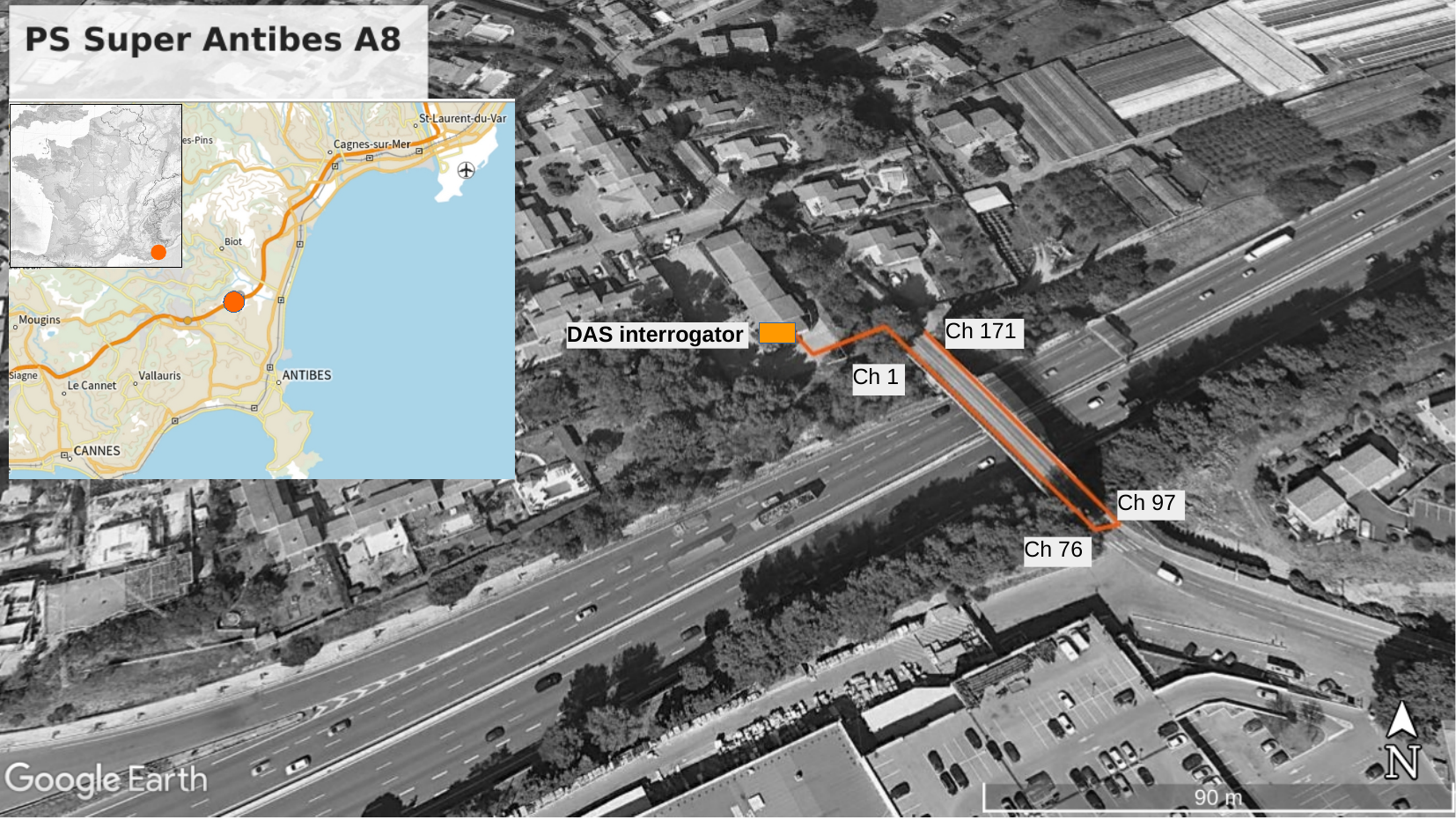}\vspace*{0.1cm}
(b)\includegraphics[width = 0.9\textwidth]{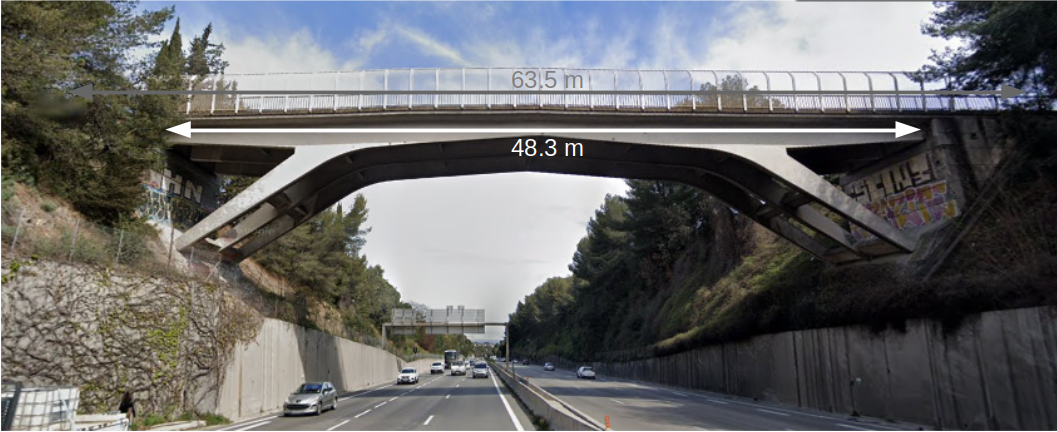}
\caption{a): Location of strut bridge over the A8 highway in southeast France. The path of the fiber is indicated in orange. The DAS interrogator was located inside a building just west of the bridge. A scale bar and a North arrow are provided on the bottom right. Imagery \textcopyright 2023 Maxar Technologies, Map data \textcopyright 2023 Google. b) view from the highway of the strut-frame bridge under study.}    
\label{fig:location}
\end{figure}

\begin{figure}[ht]
\centering    
a)\includegraphics[width=0.37\textwidth]{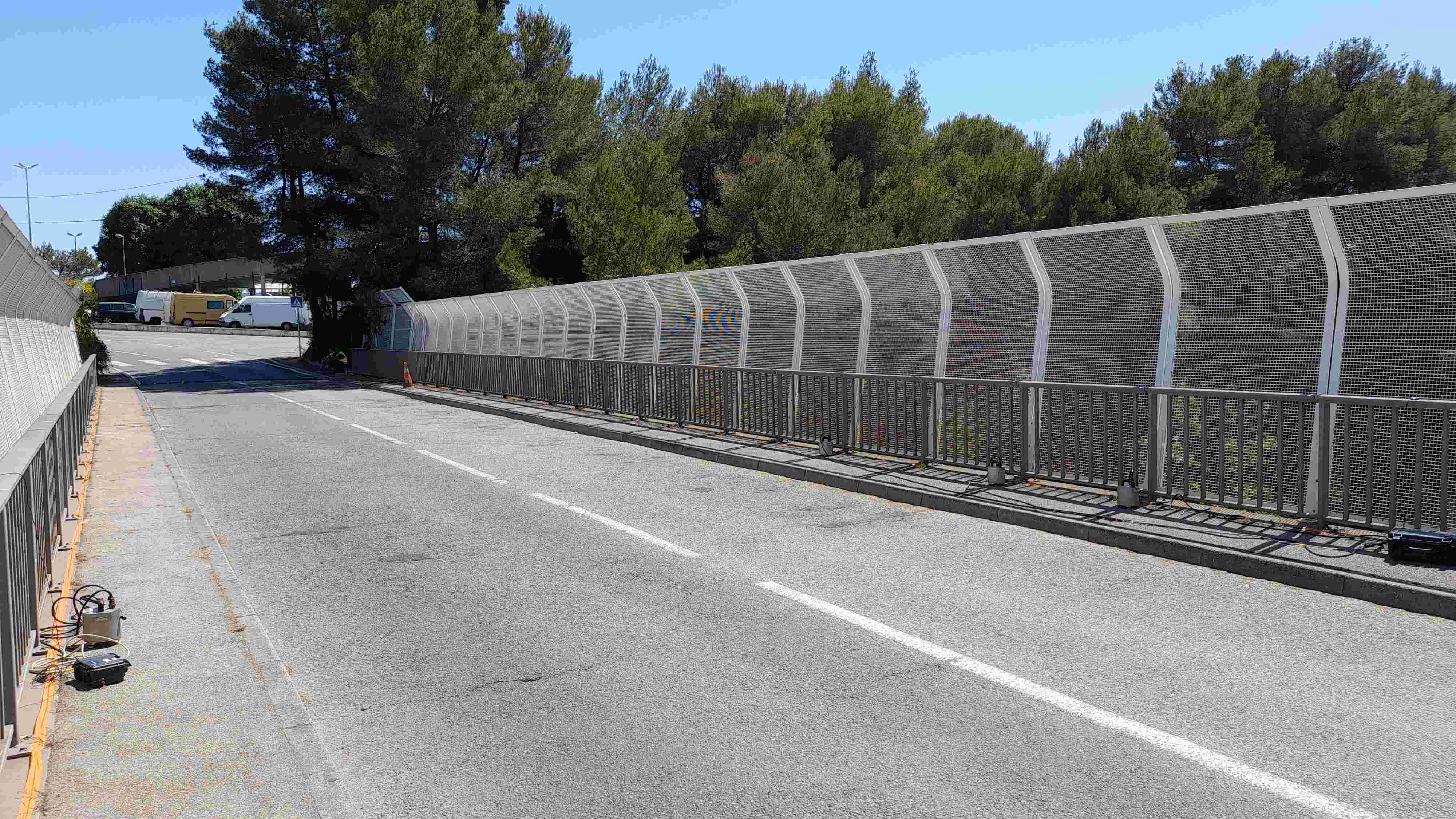}\hfill
b)\includegraphics[width=0.18\textwidth]{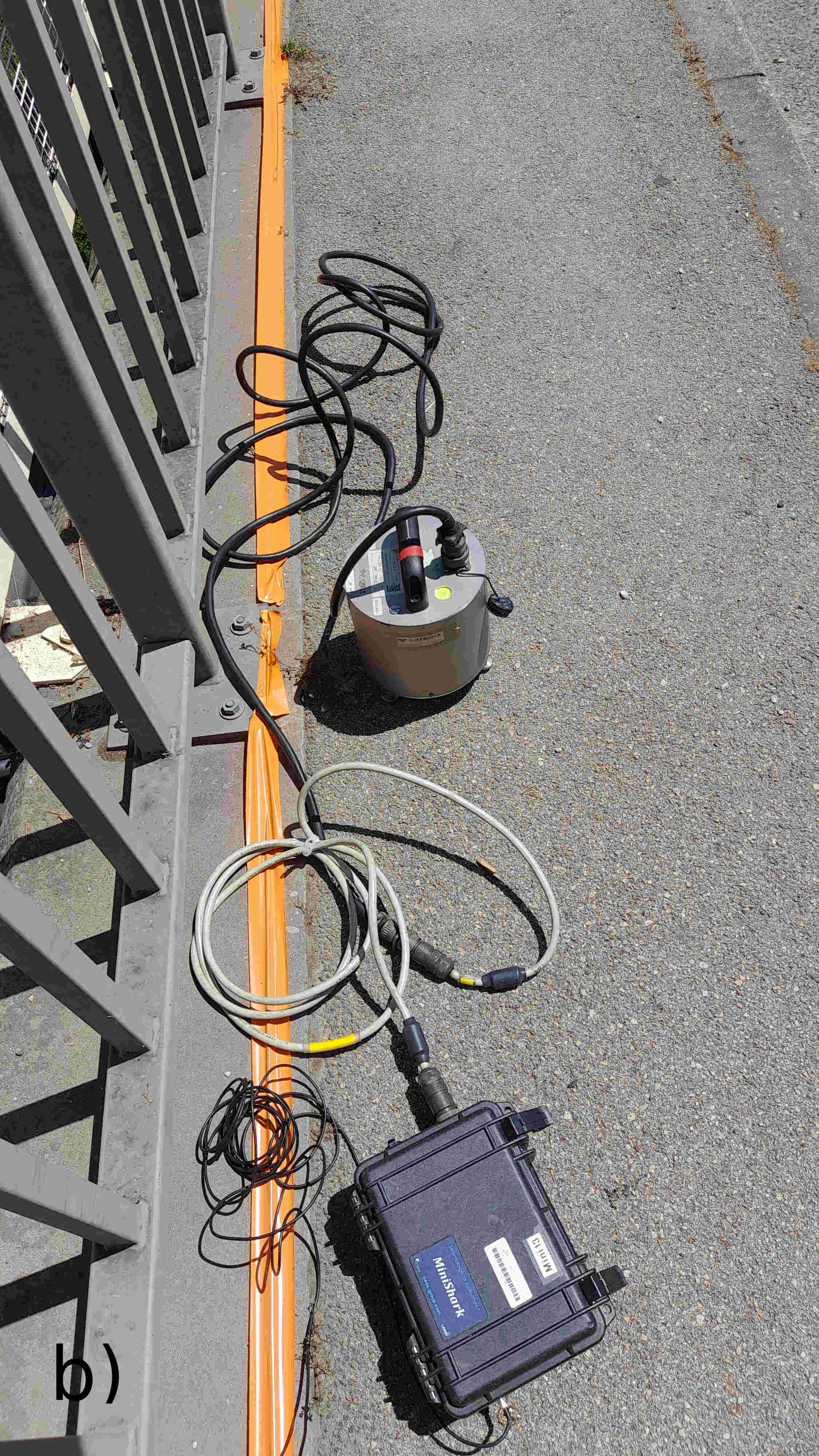}\hfill
c)\includegraphics[width=0.37\textwidth]{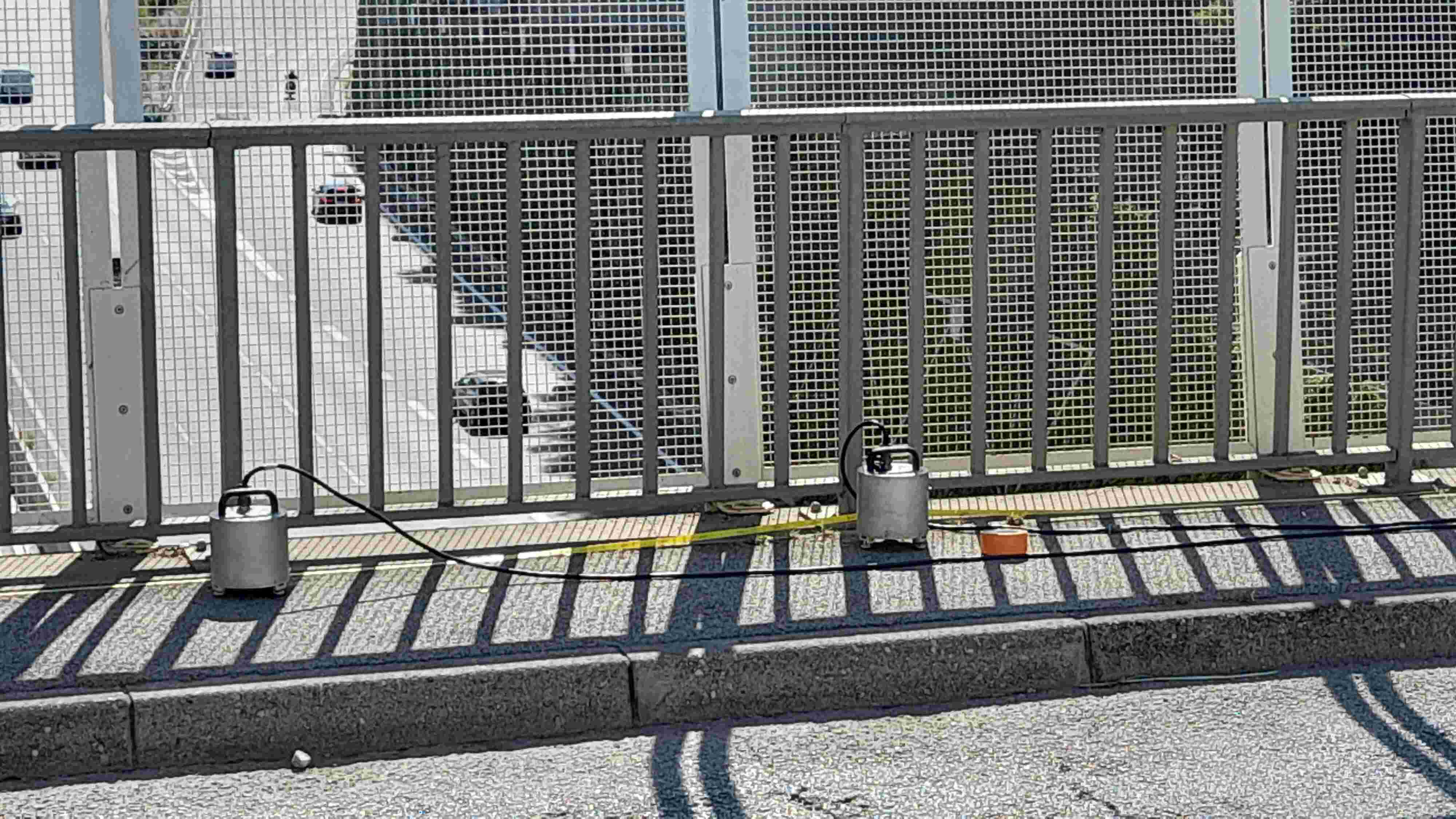}
\caption{Pictures taken during the DAS survey on June 2nd, 2022. a) general view of the bridge deck b) detail of one seismometer station and the layout of the fiber secured under the orange tape, and c) two seismometer stations side-by-side.}    
\label{fig:pictures}
\end{figure}

Two DAS experiments were carried out on a strut-frame overpass bridging the A8 highway near the city of Antibes (see location in Fig.~\ref{fig:location}). The first DAS experiment was carried out on June 2, 2022; while the second one on January 25, 2024. The strut-frame bridge under study is built on pre-stressed concrete and it is 63.5~m long and 8~m wide (see Fig \ref{fig:pictures}). The overpass bridge hosts moderate urban traffic, particularly during work hours when the DAS survey was conducted. This structure was the subject of a previous study in January 2019 \citep{perraultOperationalAnalysisTwo2019} done by the Cerema Seismic Risk team, the results of which serve as a reference for this exploratory DAS study.

For both experiments, we deployed a non-metallic tactical cable with two tight buffered single-mode optic fibers (reference BRUthough 2F from Solifos). On June 2022, the cable was deployed starting from the northern abutment, running along the western side of the deck toward the south, subsequently crossing the road some meters beyond the southern abutment of the bridge, to finally return along the eastern side of the structure. This back-and-forth deployment on each side of the bridge should in principle allow the identification of both vertical and transverse bending modes and also torsional modes of the deck. The cable was secured to the sidewalk pavement with tape as a means of providing an effective contact to the structure. The section of cable that traversed the road was protected from passing traffic by placing it in a hard rubber duct. In January 2024, on the contrary, only one side of the deck was instrumented, due to logistical constraints. During deployment, manual tap tests were performed to identify critical locations (such as the turning points of the cable) in the DAS data. In contrast to the use of 'tap test' in seismic studies, here it quite literally refers to tapping the fiber by hand for a few seconds. With the extreme sensitivity of DAS, this operation was recorded as a strong yet highly localized perturbation, which allowed us to relate the optical distance to a particular location on the bridge deck. For both surveys, data were acquired in strain-rate with an A1 interrogator of the Febus Optics company. The instrument was installed inside a building less than 100 meters from the bridge. Data were acquired at a temporal sampling rate of 8~kHz, a spatial resolution (gauge length) of 4~m, and a spatial sampling interval of 0.8~m (channel spacing). The acquisition lasted just over two hours on June 2022, generating a dataset of 230~Gb in size. On January 2024, the acquisition lasted almost 20 hours, at a slightly higher spatial sampling of 1~m, giving rise to a dataset of more than 650~Gb. It is important to note that DAS is a relative strain(-rate) measurement that does not rely on a fixed baseline. The temperature difference between the summer and winter deployments therefore has no influence on the strain measurement itself, as the baseline is arbitrarily set to zero in both surveys, and strain is therefore measured relative to zero. During DAS deployment and acquisition, two seismometers (Guralp CMG40T) connected to Minishark mobile stations were placed on the deck beside the optic fiber to act as references (see Figure \ref{fig:pictures}).
 
\subsection*{Data analysis}

\begin{figure}[t]
(a)\includegraphics[width = 0.49\textwidth]{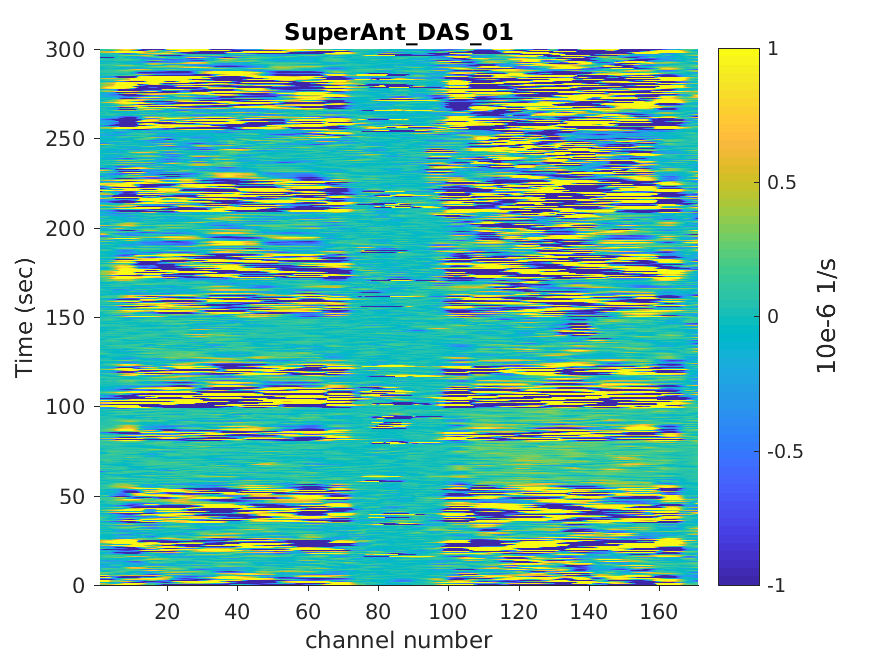}\hfill
(b)\includegraphics[width = 0.49\textwidth]{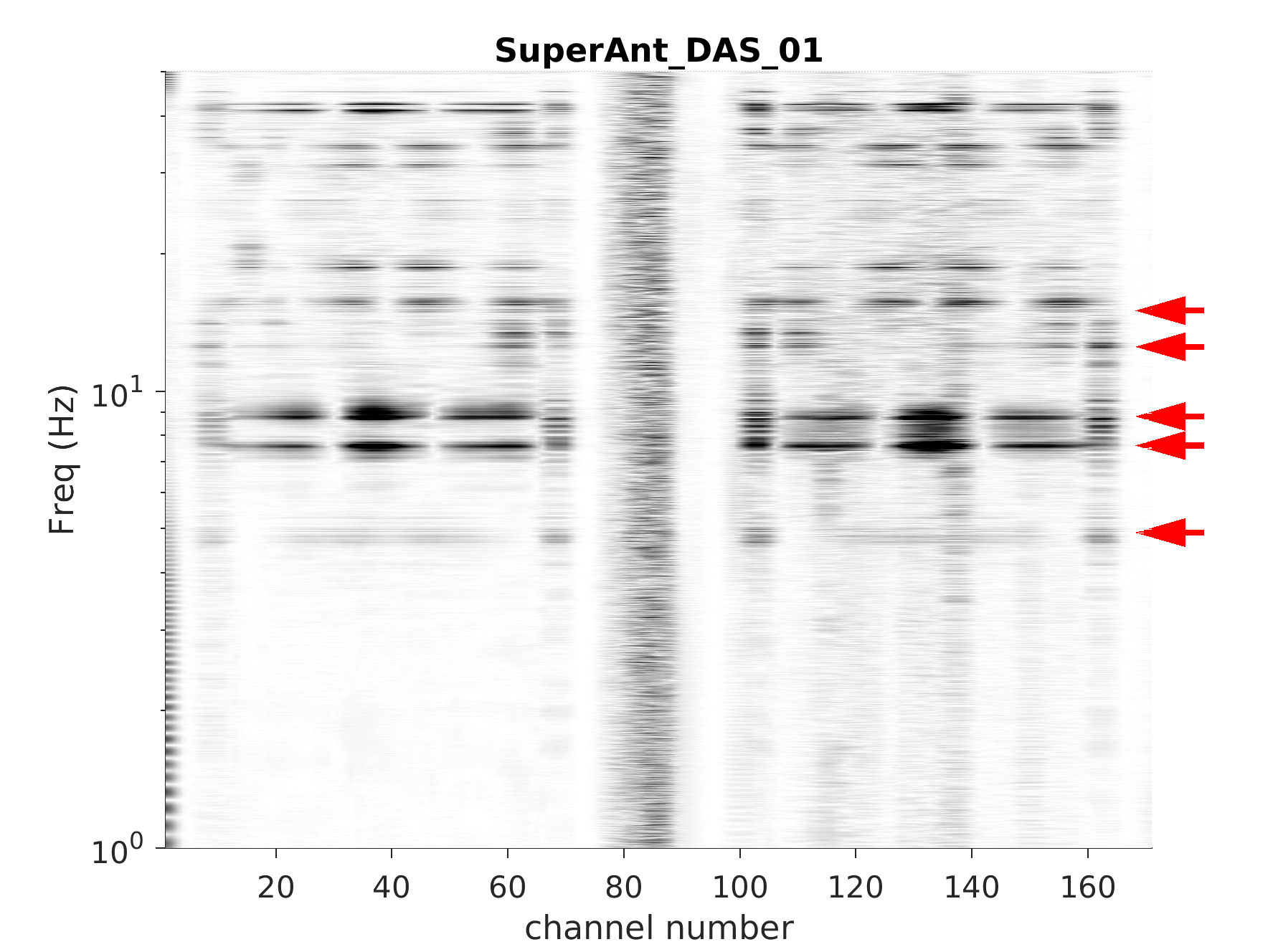}
\caption{a) extract of five minutes of strain rate data recorded by the optic fiber. b) Fourier spectra of each channel. Channels 1 to 76 correspond to the western side of the deck, while channels 97 to 171 to the eastern side of the deck.}    
\label{fig:data_spectrum}
\end{figure}

To prepare for the OMA, the DAS data was pre-processed and downsampled in time to a sampling rate of 500~Hz. The spatial sampling was 0.8 meters and the gauge length used was 4 meters. As commonly in DAS measurement, the higher the gauge length, the better the signal-to-noise ratio (SNR) \citep{deanEffectGaugeLength2017,willisGaugeLengthPulse2022}. The DAS channels that were not located on the bridge deck were removed. Subsequently, a spectral analysis was performed. Rather than converting the data into translational motion (which is an emerging practice in DAS studies in seismology \citep{vandenendeEvaluatingSeismicBeamforming2021,liorStrainGroundMotion2021,trabattoniStrainDisplacementUsing2023a}, we directly analyzed the longitudinal strain rate. A five-minute sample of ambient vibrations recorded by the fiber and their corresponding Fourier spectra is given in Fig.~\ref{fig:data_spectrum}. Channels 1 to 76 correspond to the western side of the deck, while channels 97 to 171 correspond to the eastern side. Maximum strain rates around $2 \times 10^{-5}$ to $5 \times 10^{-5}$ 1/s were recorded during the experiment while traffic passed through the bridge, although for visualization purposes, we clipped the color scale of Fig~\ref{fig:data_spectrum} to 1 microstrain. It is noticeable that the recordings of the channels that cross the road (channels 77 to 96) are not as energetic as the other ones located on the deck when there is no strong external excitation as a vehicle passing through: this difference illustrates how a simple measure as duck-taping the cable can improve the coupling with the structure. However, the passage of vehicles was recorded by these few channels as violent impulses that saturated the measurements, and therefore the signals of these channels were not usable. Fortunately, the spectra from the other channels located on the sides of the deck clearly show the presence of resonance frequencies of the bridge, identifiable as horizontal lines, i.e. similar frequencies at all the channels. In Fig~\ref{fig:data_spectrum}, resonances at frequencies 4.9 Hz, 7.6 Hz, 8.8 Hz, 12 Hz and 14.5 Hz are clearly seen (marked with red arrows in Fig~\ref{fig:data_spectrum}) and also at frequencies higher than 20 Hz. 

Based on the spectral analysis of each DAS channel, we can directly identify the modal frequencies of the bridge. Using merely several minutes of recordings, the spectral peaks emerge and the high spatial resolution of the DAS (0.8~m channel spacing) allow us to locate the nodal points (positions with zero amplitude at a particular frequency) of several modes. We also note that the quality of transient signals, such as cars and buses crossing the bridge, is insufficient for further analysis involving, for example, traffic count and characterization or traffic speed evaluation, as has been done in \cite{vandenendeDeepDeconvolutionTraffic2023,zhangAutomatedTrafficSignal2024}. Fortunately, as we will demonstrate in the next section, the DAS recordings of stationary signals along the deck are of sufficient quality to identify the normal modes of the structure.    

\section*{Results of OMA from DAS data}

\begin{figure}[ht]
\centering\includegraphics[width = 0.5\textwidth]{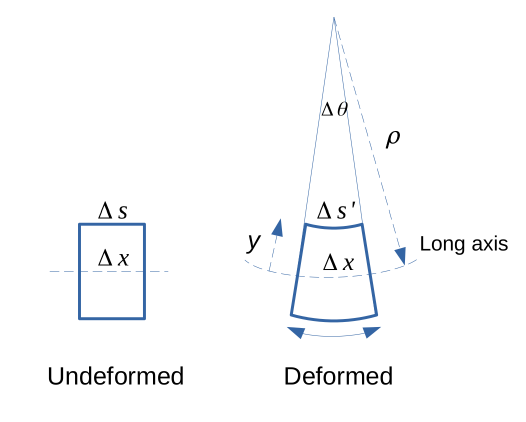}
\caption{Schema of the deformation of an infinitesimal portion of the bridge deck, when the optic fiber is located at a distance $y$ from the longitudinal neutral axis of the beam.}
\label{schema_beam}
\end{figure}

Previous works have determined the frequencies and modal shapes of bridges directly from strain time series \citep{reyndersDamageIdentificationTilff2007,anaya-diazStructuralHealthMonitoring2022,strasserStaticDynamicBridge2023}. This has the advantage of avoiding the cumbersome spatial integration step to get displacement (or velocity) time series, using strain directly for OMA. Following classical beam theory, the longitudinal normal strain ($\epsilon_l$) of a continuous beam subjected to flexural forces is directly proportional to the distance to the longitudinal (neutral) axis of the beam ($y$) \citep{landauTheorieElasticiteTheory1953}, that is $\epsilon_l\,\sim \,-y/\rho$, where $\rho$ is the radius of curvature (see Figure~\ref{schema_beam}). 
As a first-order approximation, the bridge deck can be assimilated to a horizontal beam simply supported at both ends. Considering an incremental element of the beam (Fig.~\ref{schema_beam}), it can be easily shown that any line segment $\Delta s$ located at an arbitrary distance $y$ from
the neutral axis will elongate or contract and become $\Delta s'$ after deformation. Then, by
definition, the normal strain along the longitudinal direction is determined as $\epsilon_l=(\Delta s' - \Delta s) / \Delta s$. This can also be expressed as $\epsilon_l\sim((\rho-y)\Delta\theta - \rho\Delta \theta) / \rho\,\Delta\theta = -y/\rho$. As the optic fiber cable lies at some distance from the neutral axis of the beam, this longitudinal deformation will occur cyclically during each time cycle of the corresponding normal mode.

In fact, as we can see from Fig.~\ref{fig:data_spectrum}, the vertical bending modes, i.e. with deformation mainly in the vertical direction, are the ones that present the spectral peaks with the highest amplitudes in the DAS data, sensible only to longitudinal strains (i.e. in the cable direction). 

Similarly, if we consider the flexural deformation in the transverse direction of the bridge, the longitudinal strain will also be proportional to this orthogonal deformation of the structure (now in the horizontal transverse direction). As has been observed by \cite{rodetUrbanDarkFiber2025}, the deck motions in any of the three spatial directions suffice to elongate/contract the optic fiber and therefore generate a signal recorded by DAS.   

\begin{figure}[ht]
\center{\footnotesize WINTER 2019 (AVT) \hspace*{1.4cm} SUMMER 2022 (DAS) \hspace*{1.4cm} WINTER 2024 (DAS)}\\\vspace*{0.3cm}
\includegraphics[width = 0.33\textwidth]{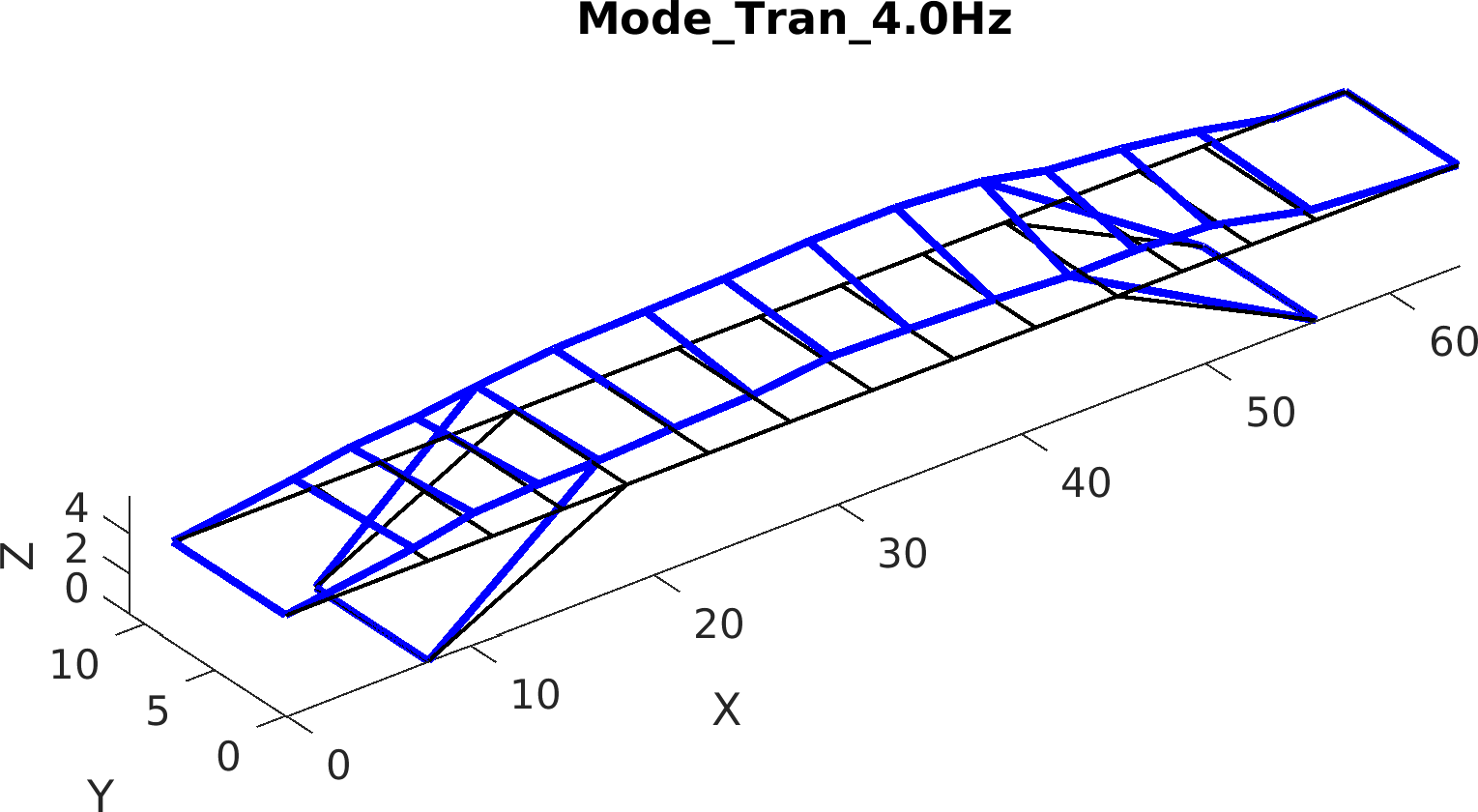}\hfill
\includegraphics[width = 0.33\textwidth]{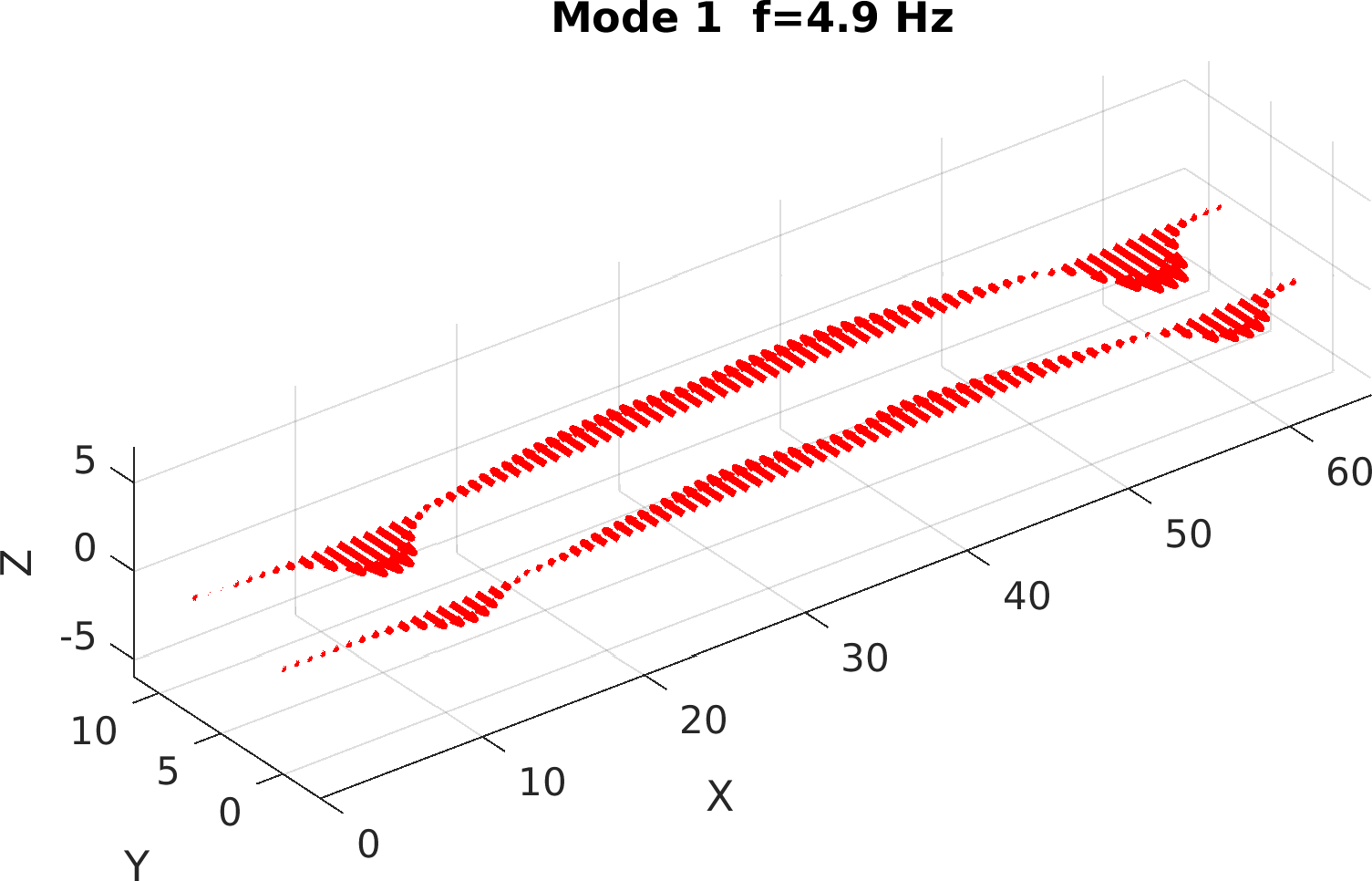}\hfill
\includegraphics[width = 0.33\textwidth]{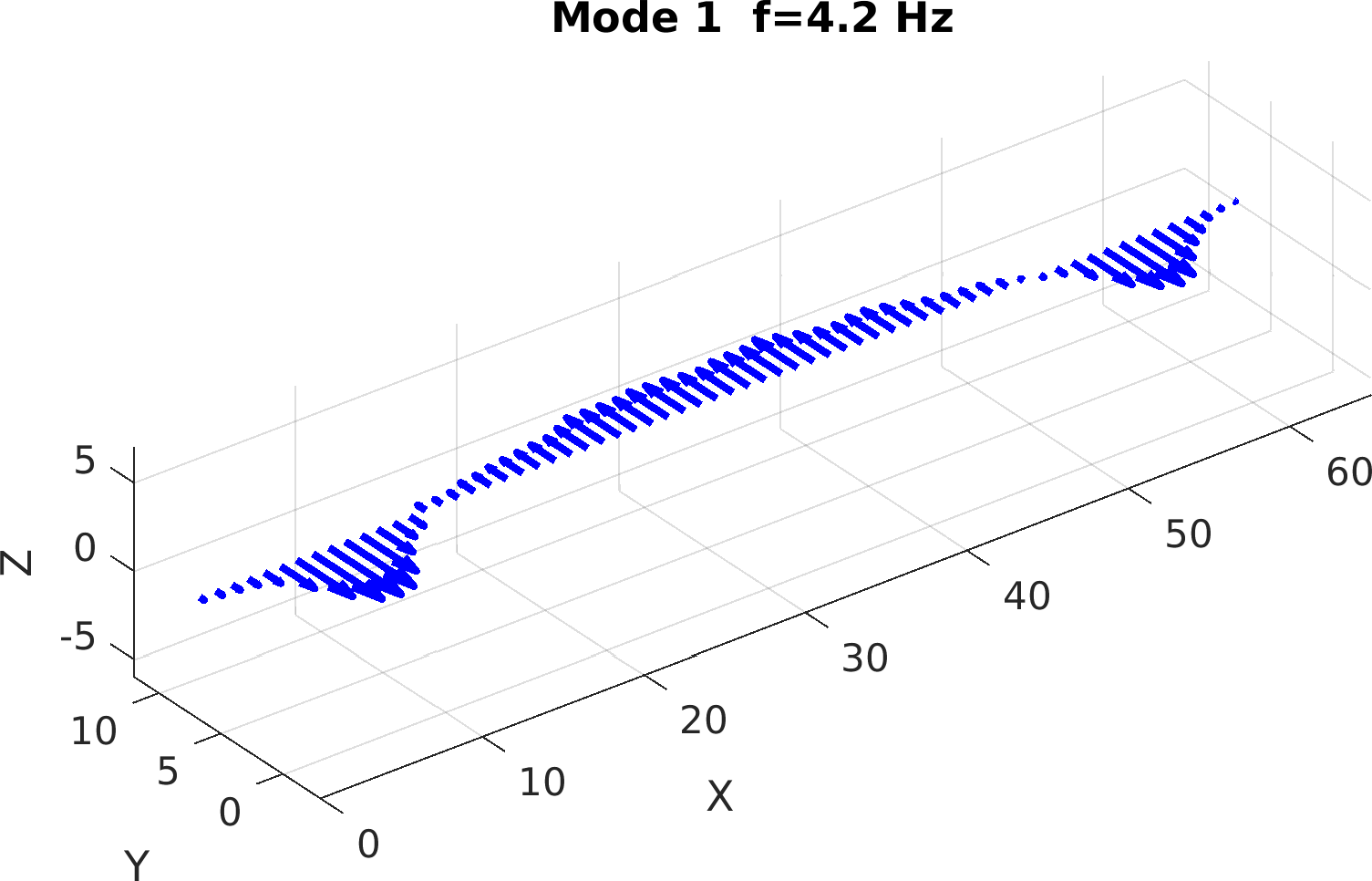}
\includegraphics[width = 0.33\textwidth]{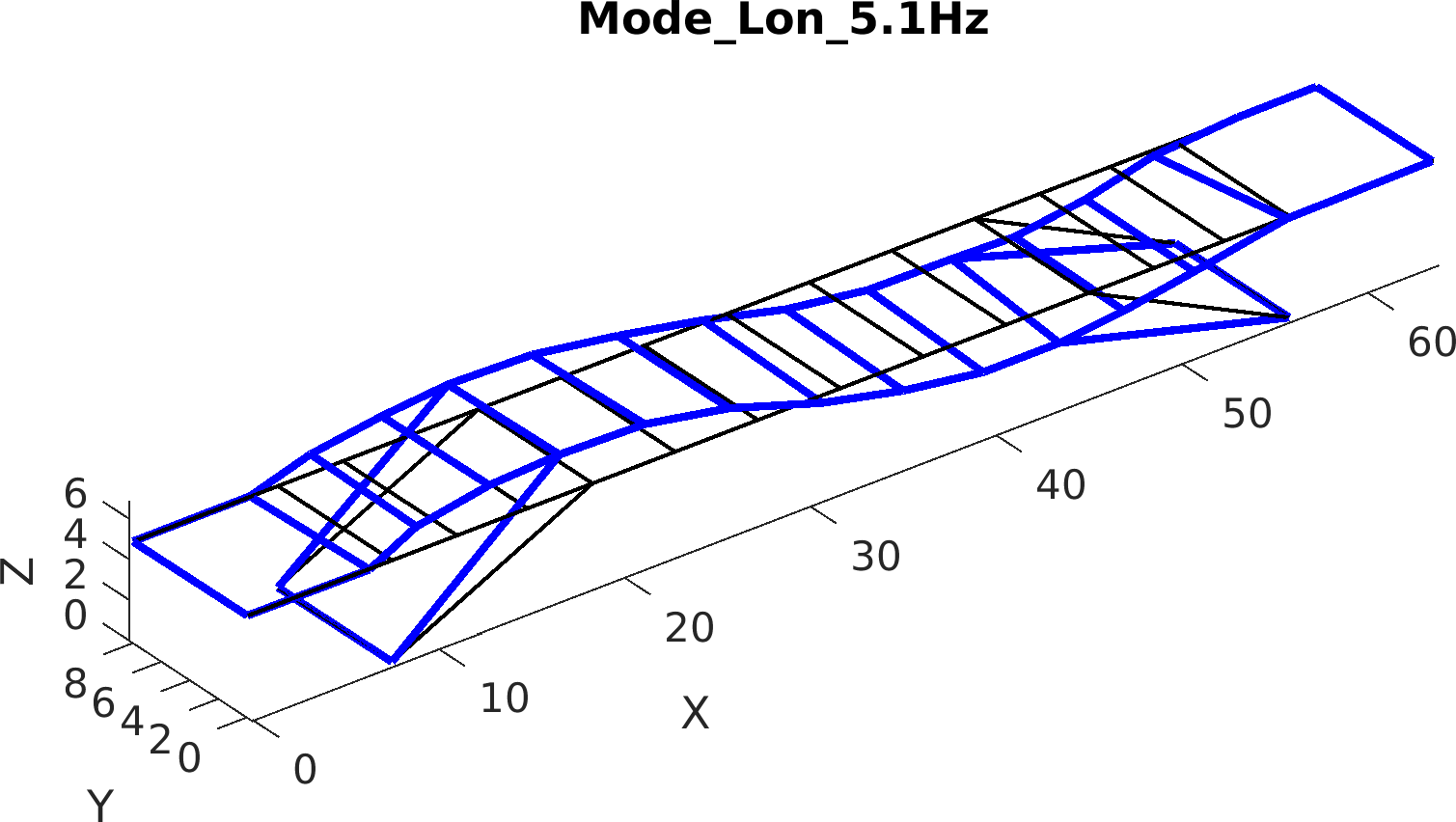}
\includegraphics[width = 0.33\textwidth]{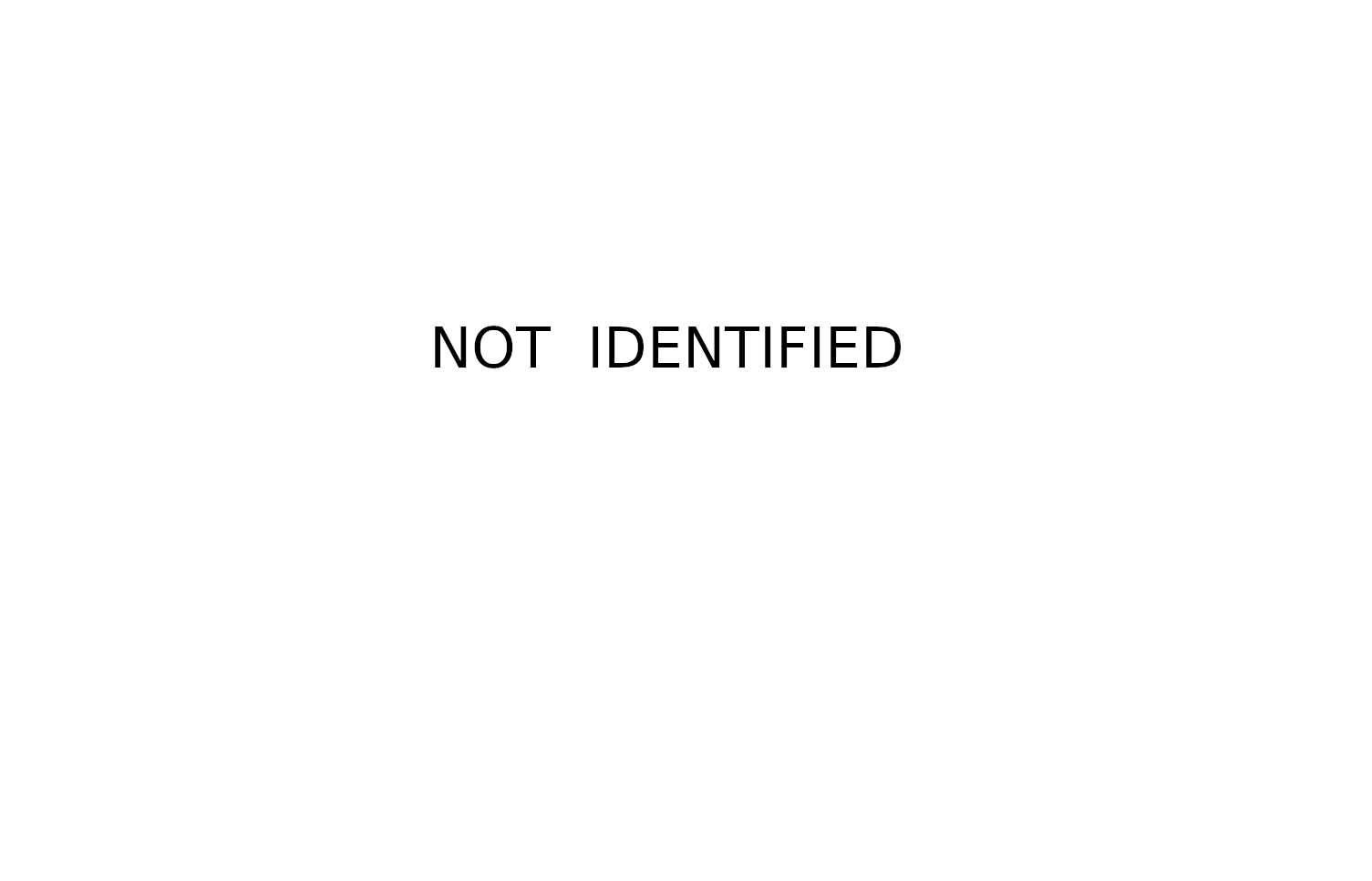}\hfill
\includegraphics[width = 0.33\textwidth]{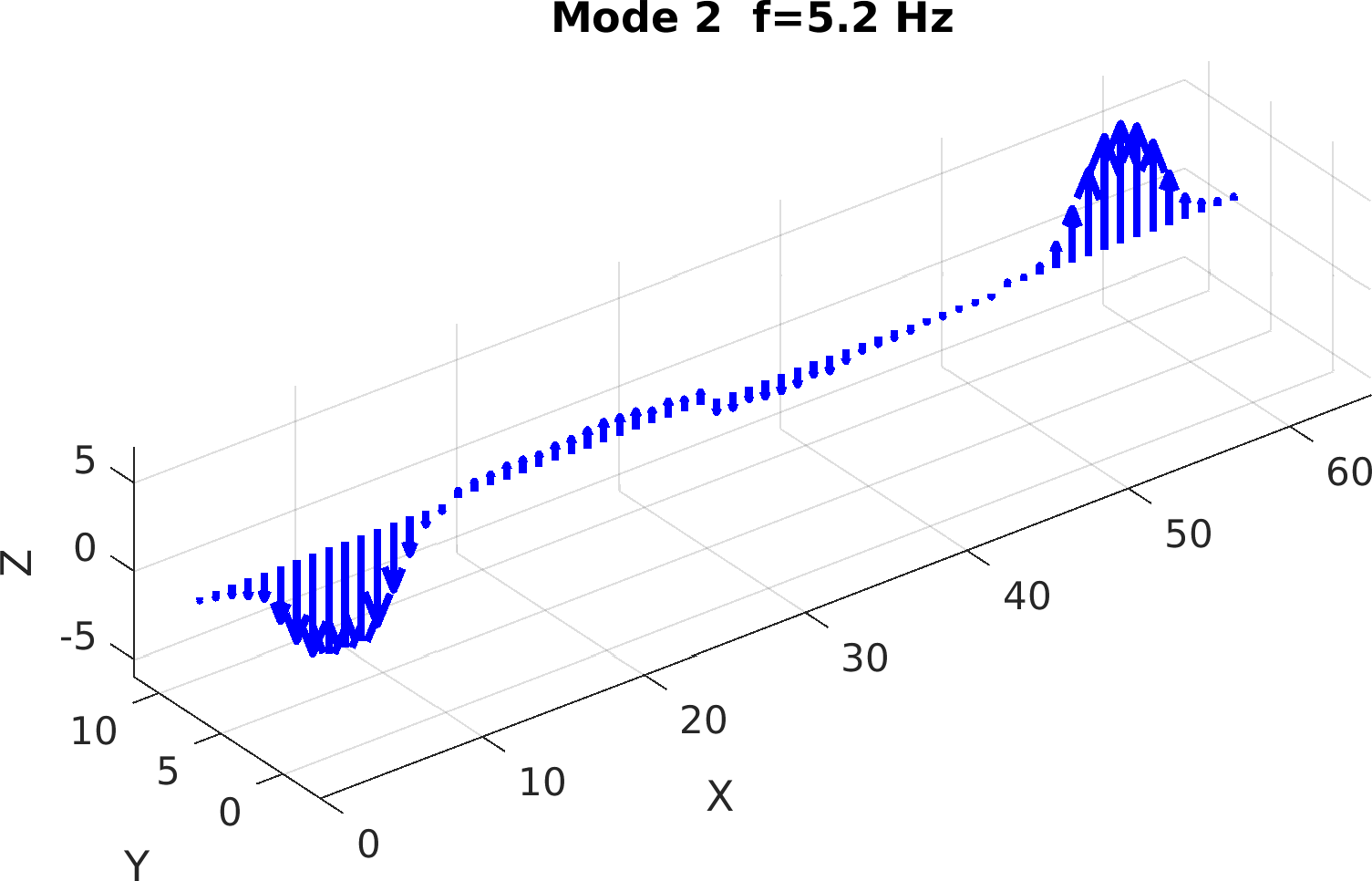}
\includegraphics[width = 0.33\textwidth]{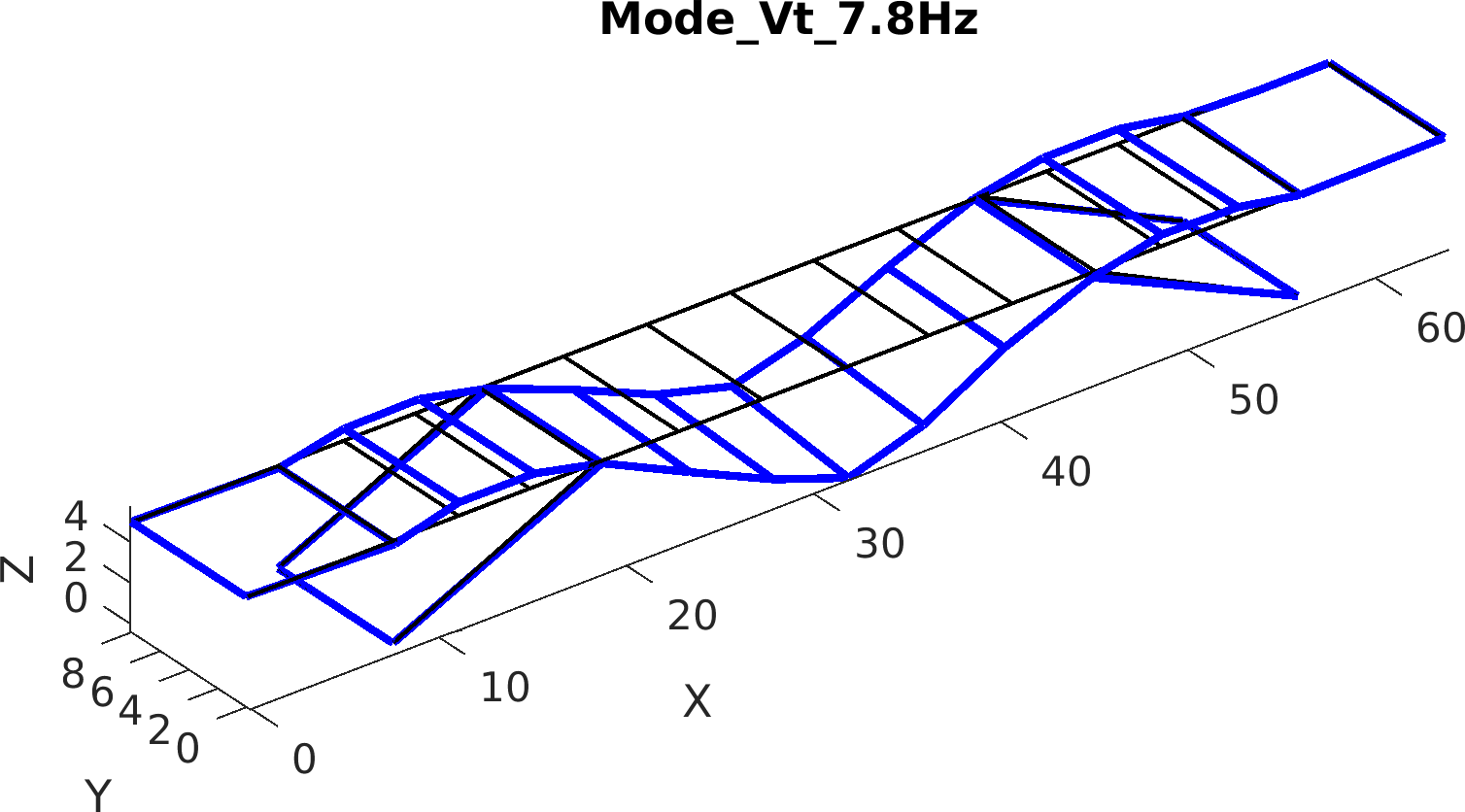}\hfill
\includegraphics[width = 0.33\textwidth]{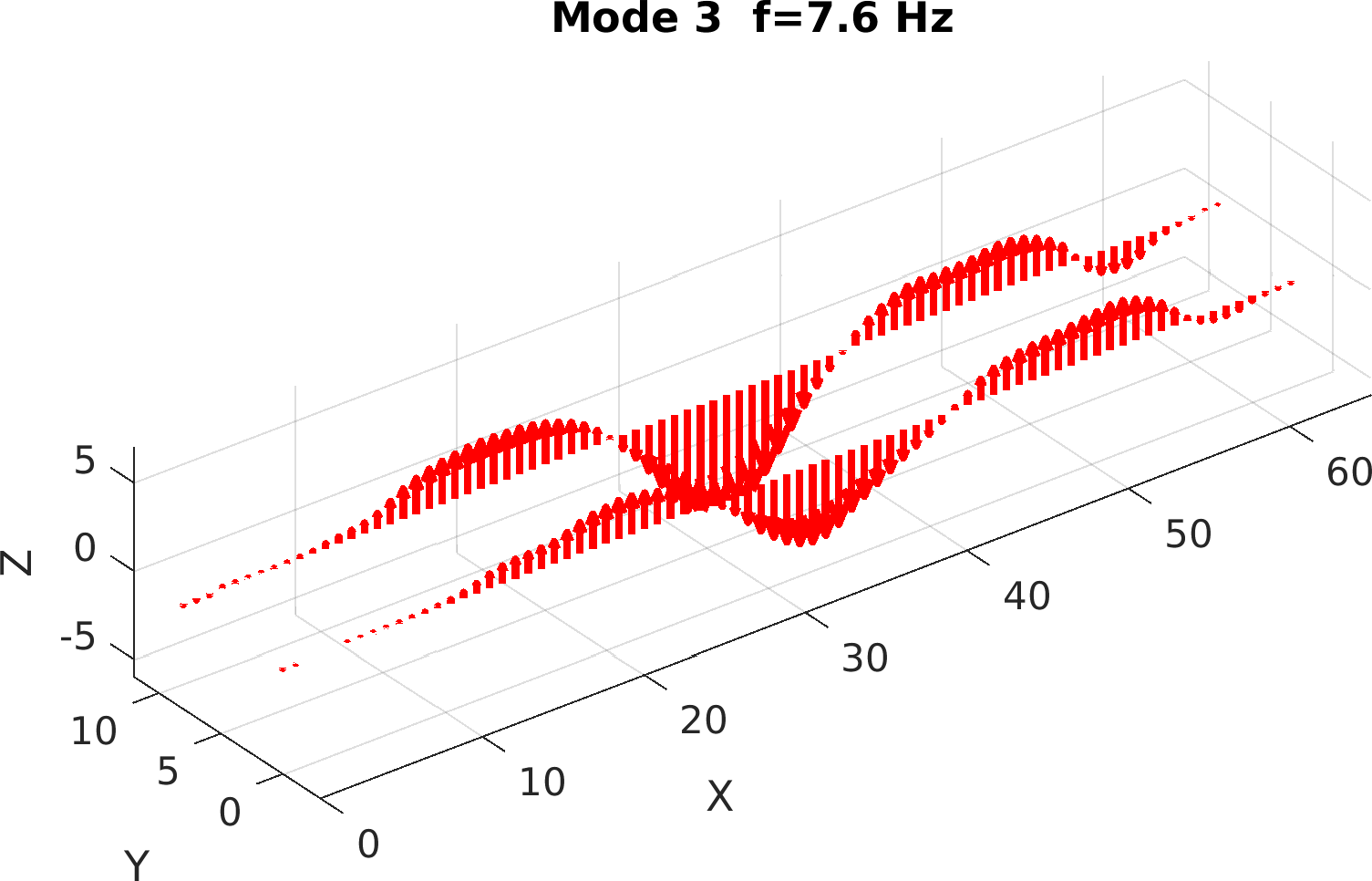}\hfill
\includegraphics[width = 0.33\textwidth]{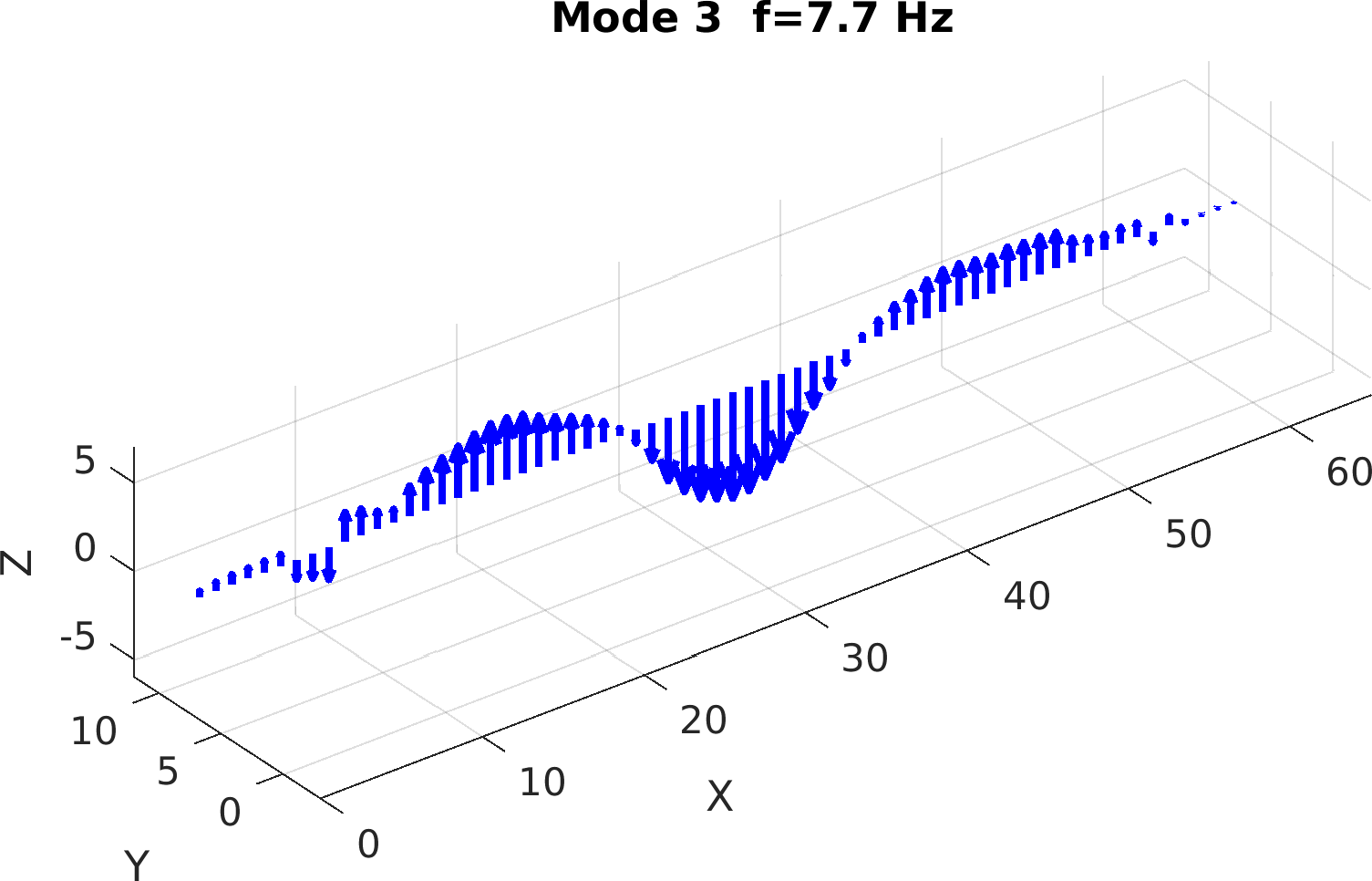}
\includegraphics[width = 0.33\textwidth]{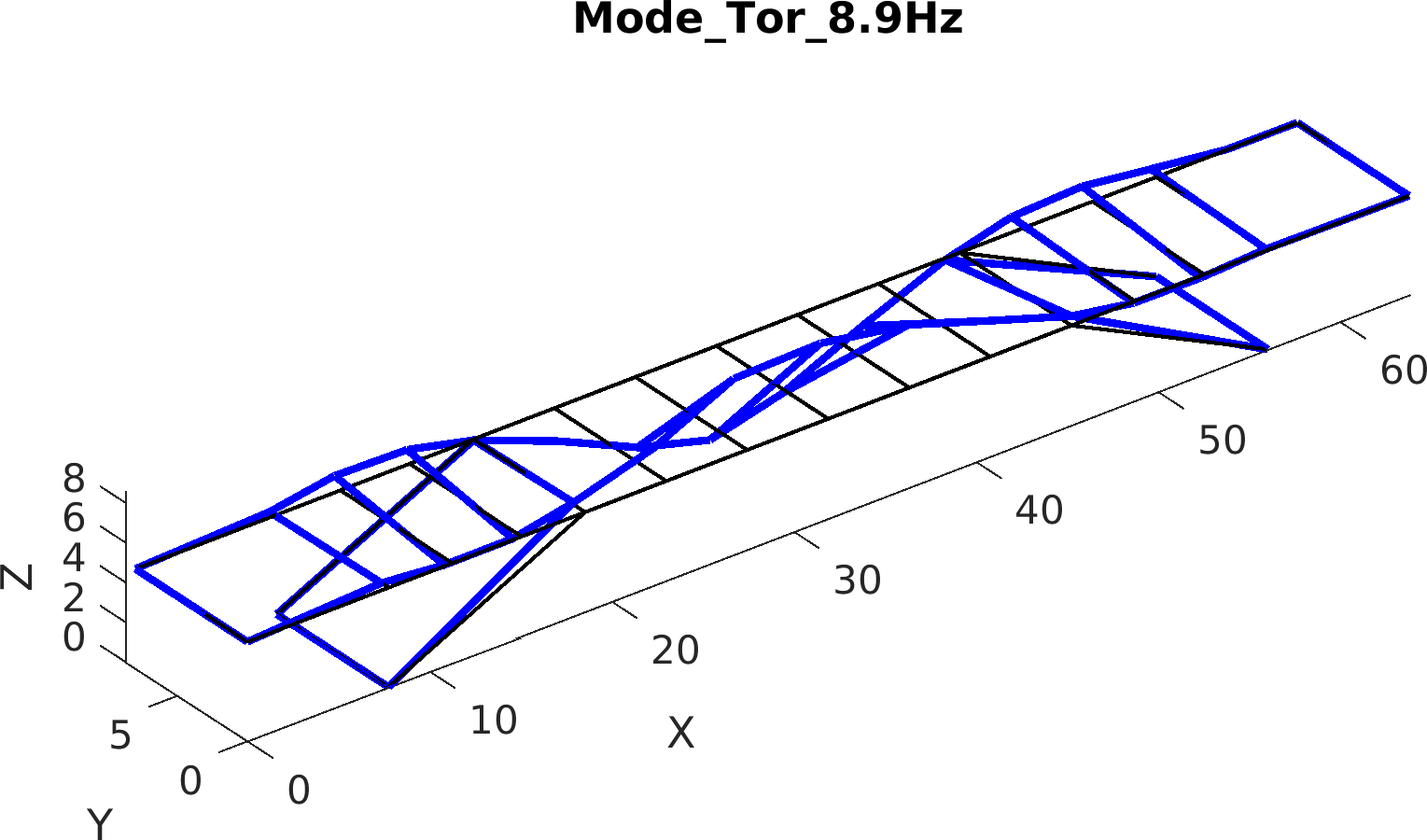}
\includegraphics[width = 0.33\textwidth]{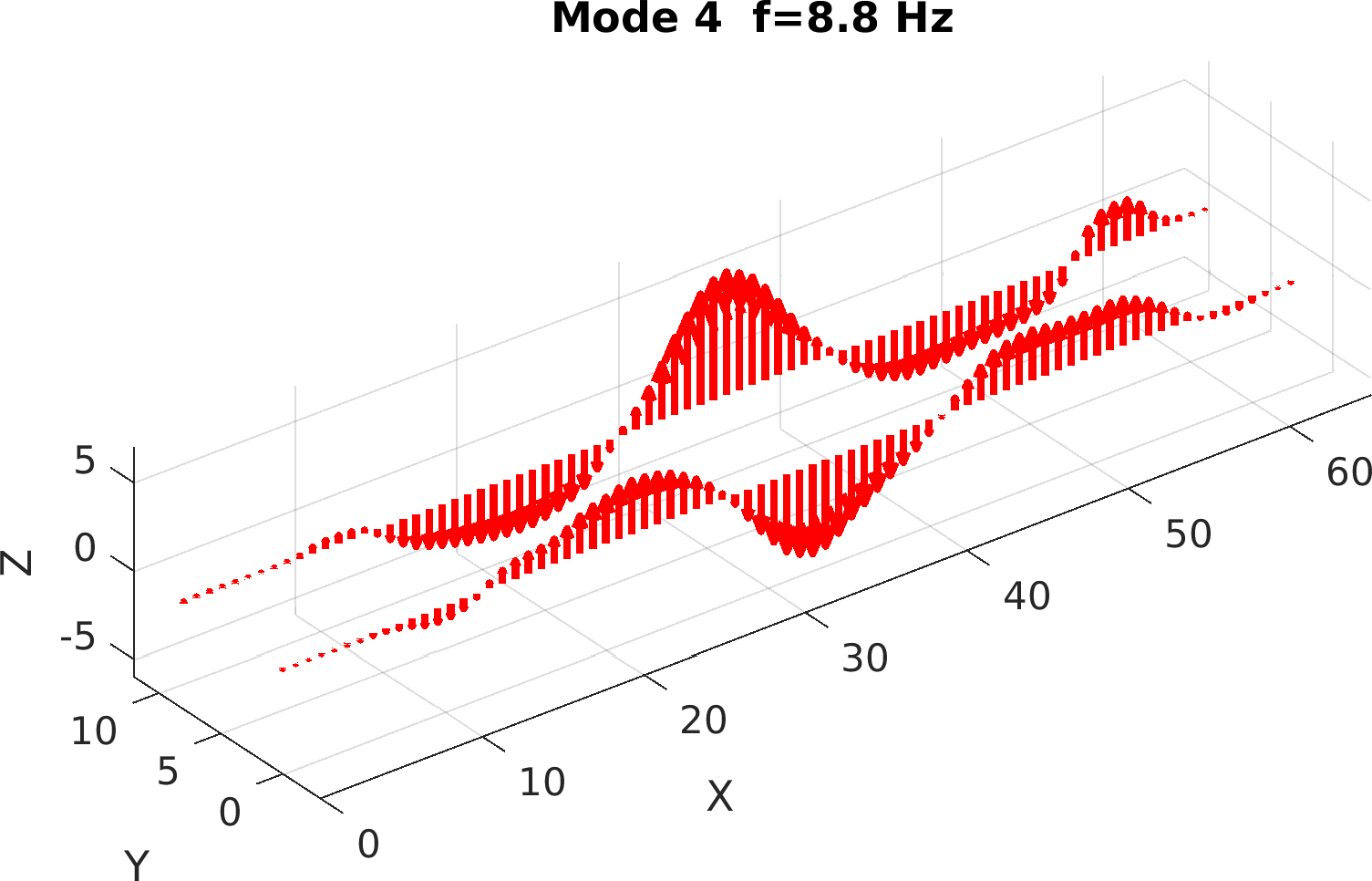}\hfill
\includegraphics[width = 0.33\textwidth]{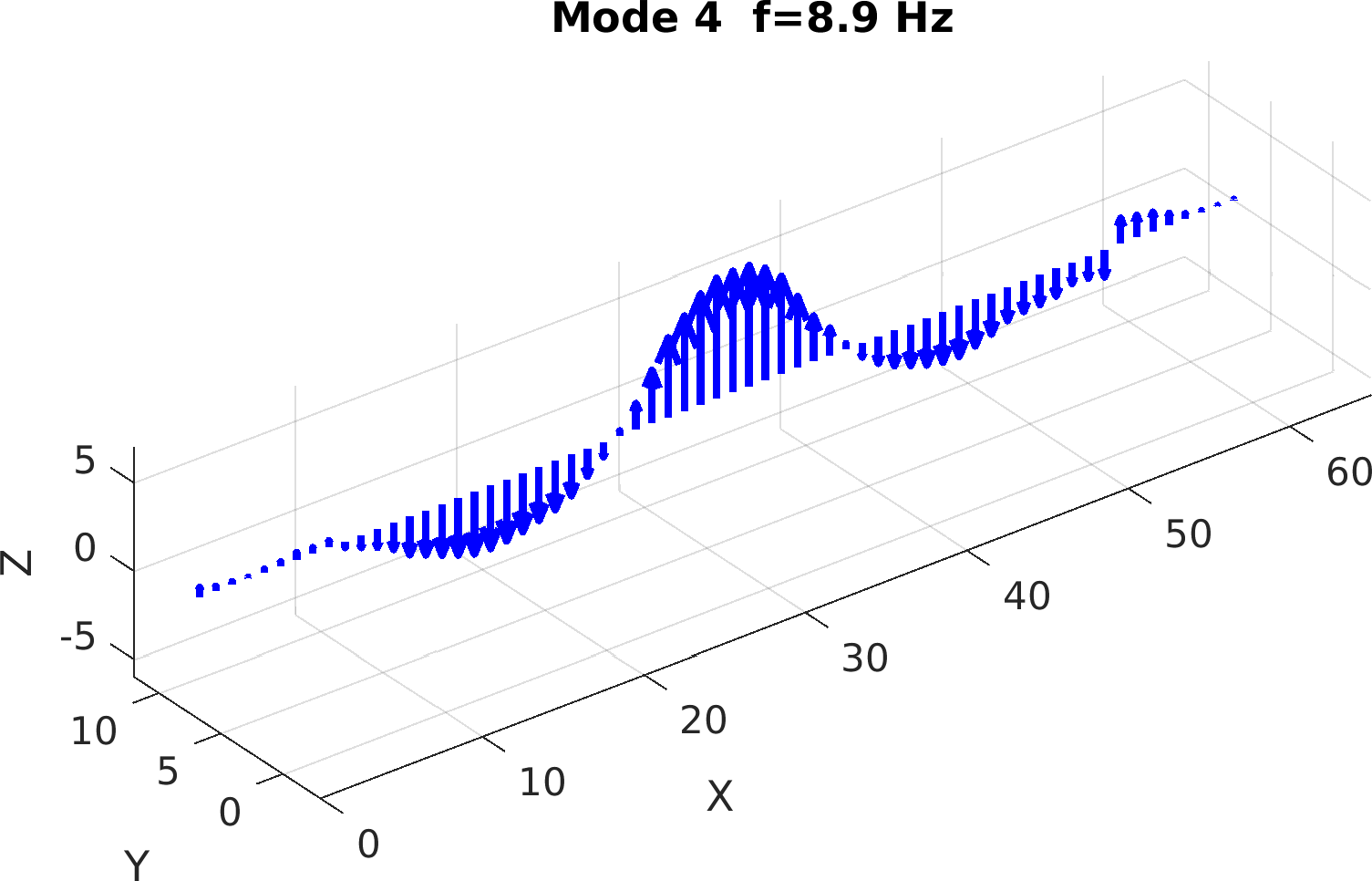}
\caption{Modal shapes of the strut-frame bridge based on seismometer data of 2019. a) transverse mode b) longitudinal mode c) 1st vertical mode d) 1st torsional mode. e), f) and g) DAS based modal shapes of the same modes (summer 2022 survey), h) i) j) and k) DAS based modal shapes of the same modes (winter 2024 survey).}    
\label{fig:modes_MP}
\end{figure}

Before presenting the OMA analysis of the DAS data, it is useful to summarize here the results of a previous study \citep{perraultOperationalAnalysisTwo2019} where the OMA analysis of the same bridge was carried out by means of the Frequency Domain Decomposition (FDD) technique \citep{brinckerModalIdentificationOutputonly2001} using conventional seismometer arrays located on the bridge deck. It is important to note that this survey was conducted during the winter of 2019, contrary to the DAS experiment done during the summer of 2022 : we will come back to this difference in the analysis of the results. The 2019 study was carried out using six CMG40T velocimeters synchronized by a single (multichannel) recording station (CityShark-v2 Leas). Several seismometer set-ups were deployed to cover and sufficiently densify the measurement locations on the deck (total of 31 different locations). After reprocessing of the ambient vibration data for the present work, the results are summarized in Fig~\ref{fig:modes_MP}. The first transverse mode is found at 4 Hz, the first longitudinal mode at 5.1 Hz, showing also an important vertical buckling due to the strut-frame and the abutments at both sides of the bridge deck. Finally, two well-expressed vertical modes appear at 7.8 Hz and 8.8 Hz. The latter corresponds to the first torsional mode of the deck (anti-symmetric with respect to the longitudinal axis of the bridge). More details of this survey and the complete OMA analysis can be found in \cite{perraultOperationalAnalysisTwo2019}.       

\begin{figure}[t]
\centering
\includegraphics[width = 0.7\textwidth]{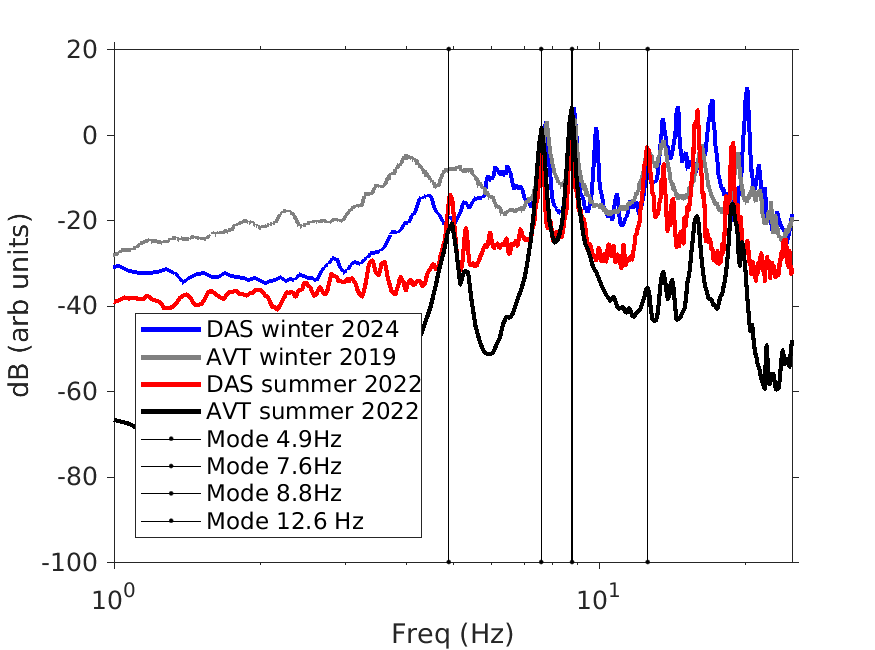}
\caption{Comparison of the first singular values of several datasets: ambient vibration testing (AVT) in the winter of 2019, AVT in the summer of 2022, and DAS in the summer of 2022 and winter of 2024. In vertical solid lines, the first four normal modes identified in 2022 are indicated.}    
\label{fig:svds}
\end{figure}

Returning to the present study using DAS data to identify the normal modes of the bridge, strain-rate vibration traces are also analyzed using the FDD technique following \cite{brinckerModalIdentificationOutputonly2001}. First, data of each channel are cut by windows of 20.48 seconds time duration (spectral resolution close to 0.05 Hz). Then, after detrending and apodisation of the signals, a bandpass filter (Butterworth 2nd order) between 0.5 Hz and 50 Hz is applied. The Cross-Power Spectral Density (CPSD) matrix is constructed as a function of frequency, and its Singular Value Decomposition is calculated. It should be noted that, at the same time the DAS data was acquired, two seismometers (similar to the ones used in 2019) have also been placed on the bridge deck to serve as reference for the vibration measurements in 2022. The first singular value for DAS and seismometer data (2019 and 2022) is shown as a function of frequency in Fig~\ref{fig:svds}. Note that the vertical scale in Fig~\ref{fig:svds} is arbitrary, we have scaled the singular values amplitudes to coincide around 8 Hz. First, the similarity of the spectral peaks of the DAS and the 2022 Ambient Vibration Testing (AVT) using seismometer data is remarkable. This result was expected, as the strain rate is known to be proportional to the translational acceleration \citep{liorStrainGroundMotion2021}. The SNR for the DAS data is inferior to that of the medium-band seismometers but still sufficient to identify the peaks. When comparing the AVT data from both winter and summer surveys, we realize that the vertical modes (7.7 Hz and 8.8 Hz) are quite consistent between the different years although we observe a slight increase ($< 2\% $) in frequencies in winter (7.7 Hz and 8.8 Hz) with respect to summer (7.6 Hz and 8.7 Hz). The same can be said for higher modes at higher frequencies. In contrast, the first peak in the 2019 AVT at 4 Hz has been shifted relative to the 2022 AVT to 4.9 Hz, and the longitudinal mode frequency peak at 5.1 Hz in 2019 is missing (or hidden by another frequency peak) in both the 2022 AVT and DAS datasets recorded during the summer.

\begin{figure}[ht]
\center{\footnotesize WINTER 2019 (AVT) \hspace*{1.4cm} SUMMER 2022 (DAS) \hspace*{1.4cm} WINTER 2024 (DAS)}\\\vspace*{0.3cm}
\includegraphics[width = 0.33\textwidth]{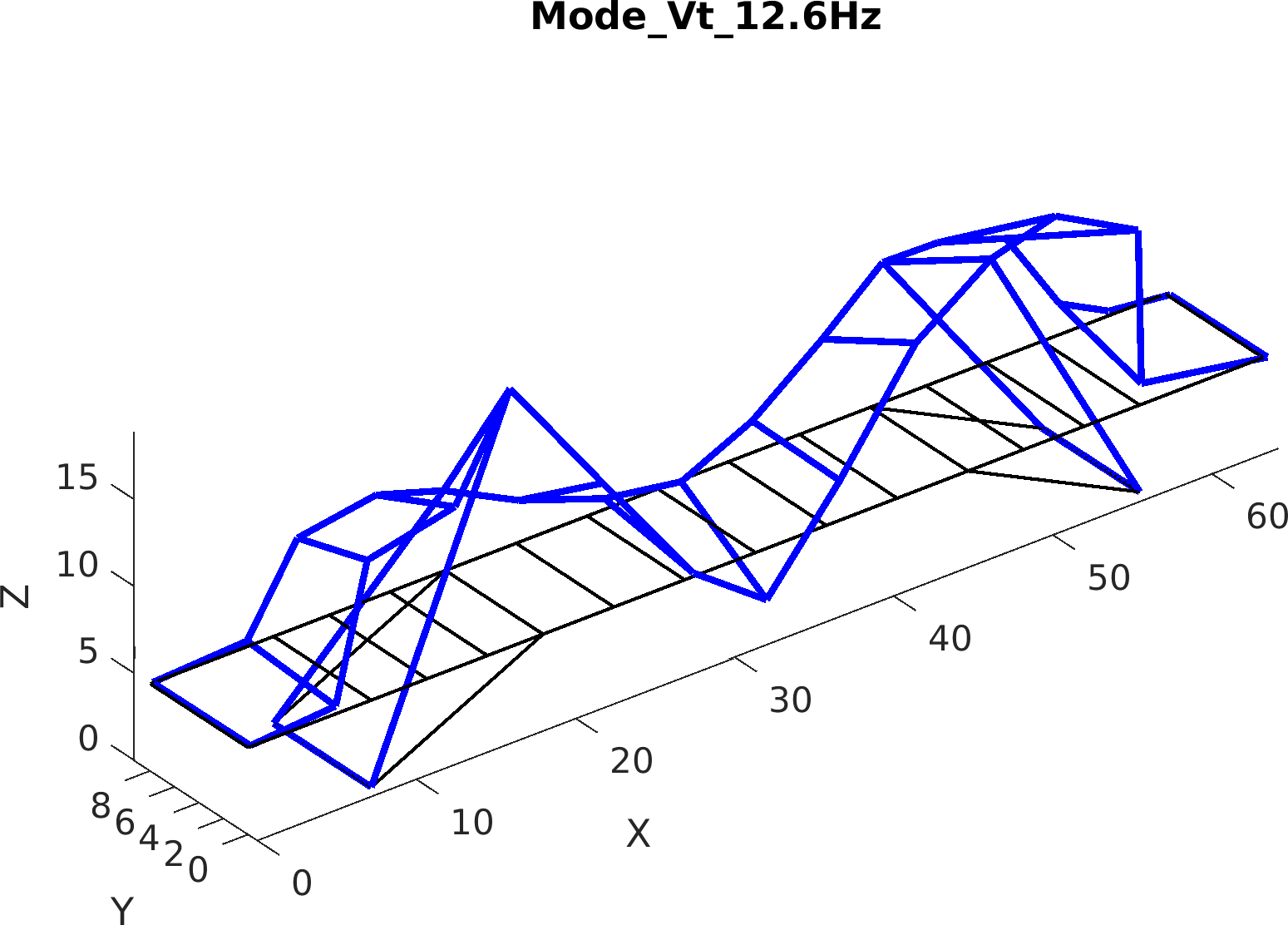}\hfill
\includegraphics[width = 0.33\textwidth]{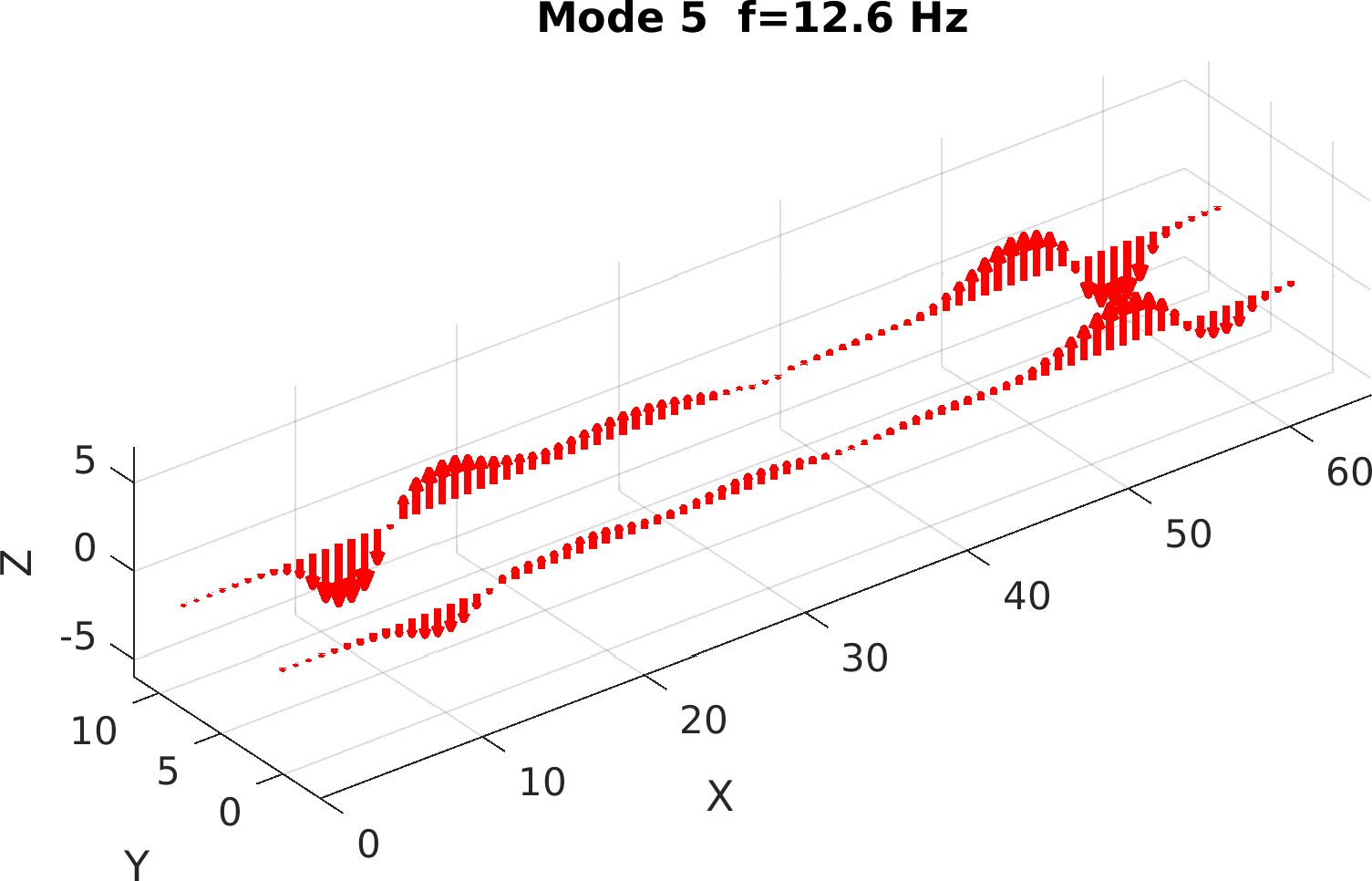}\hfill
\includegraphics[width = 0.33\textwidth]{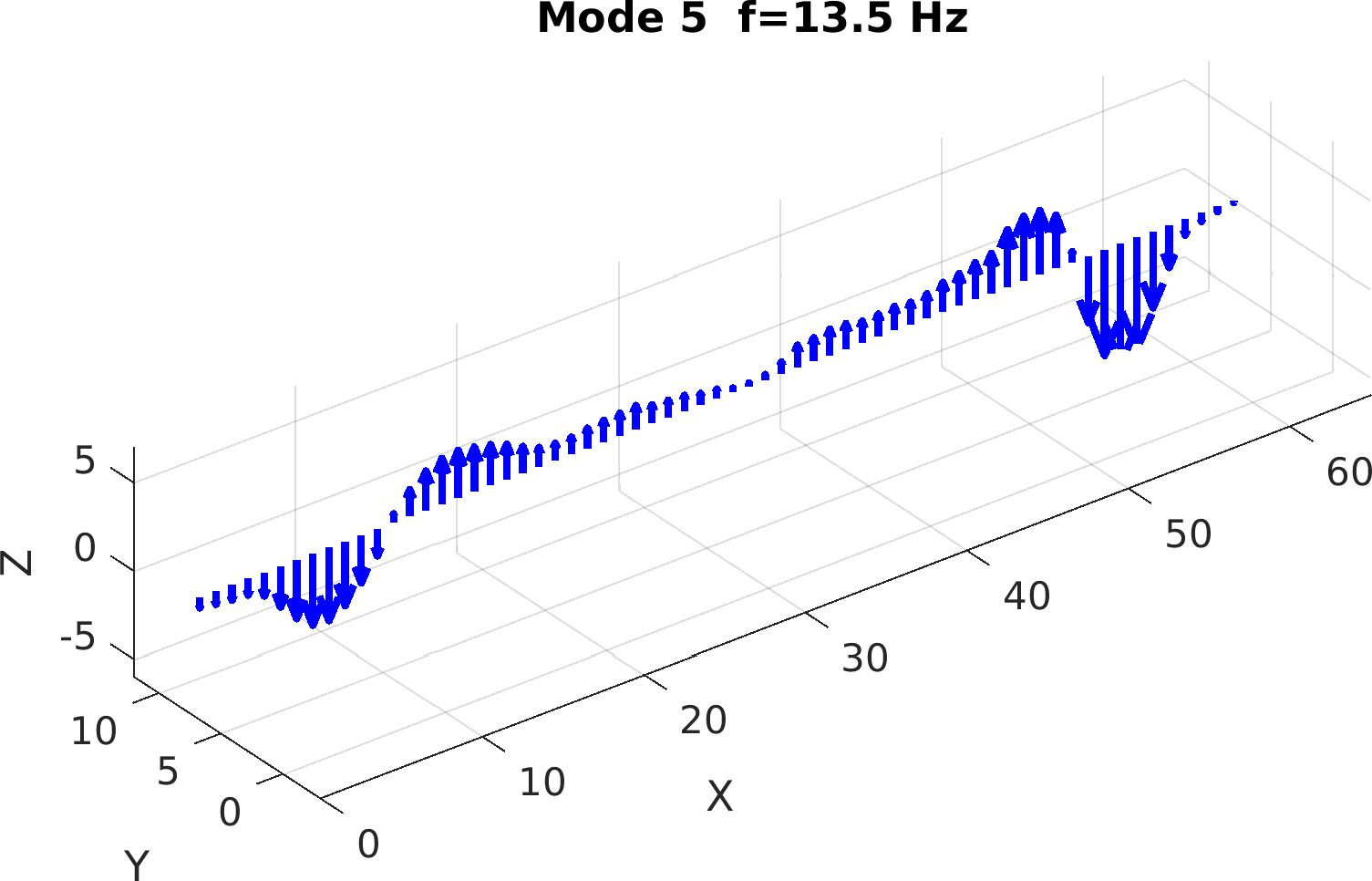}
\includegraphics[width = 0.33\textwidth]{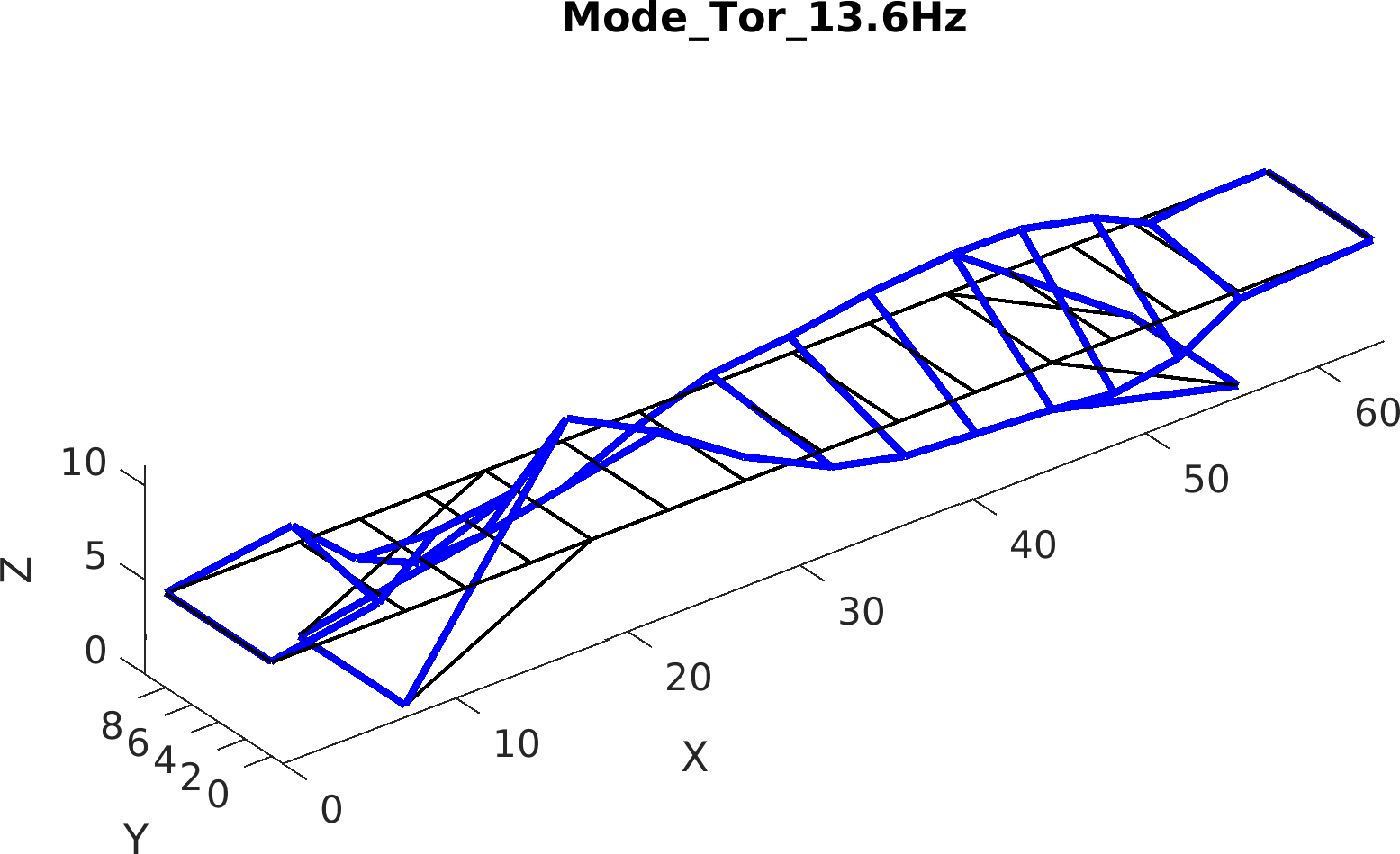}\hfill
\includegraphics[width = 0.33\textwidth]{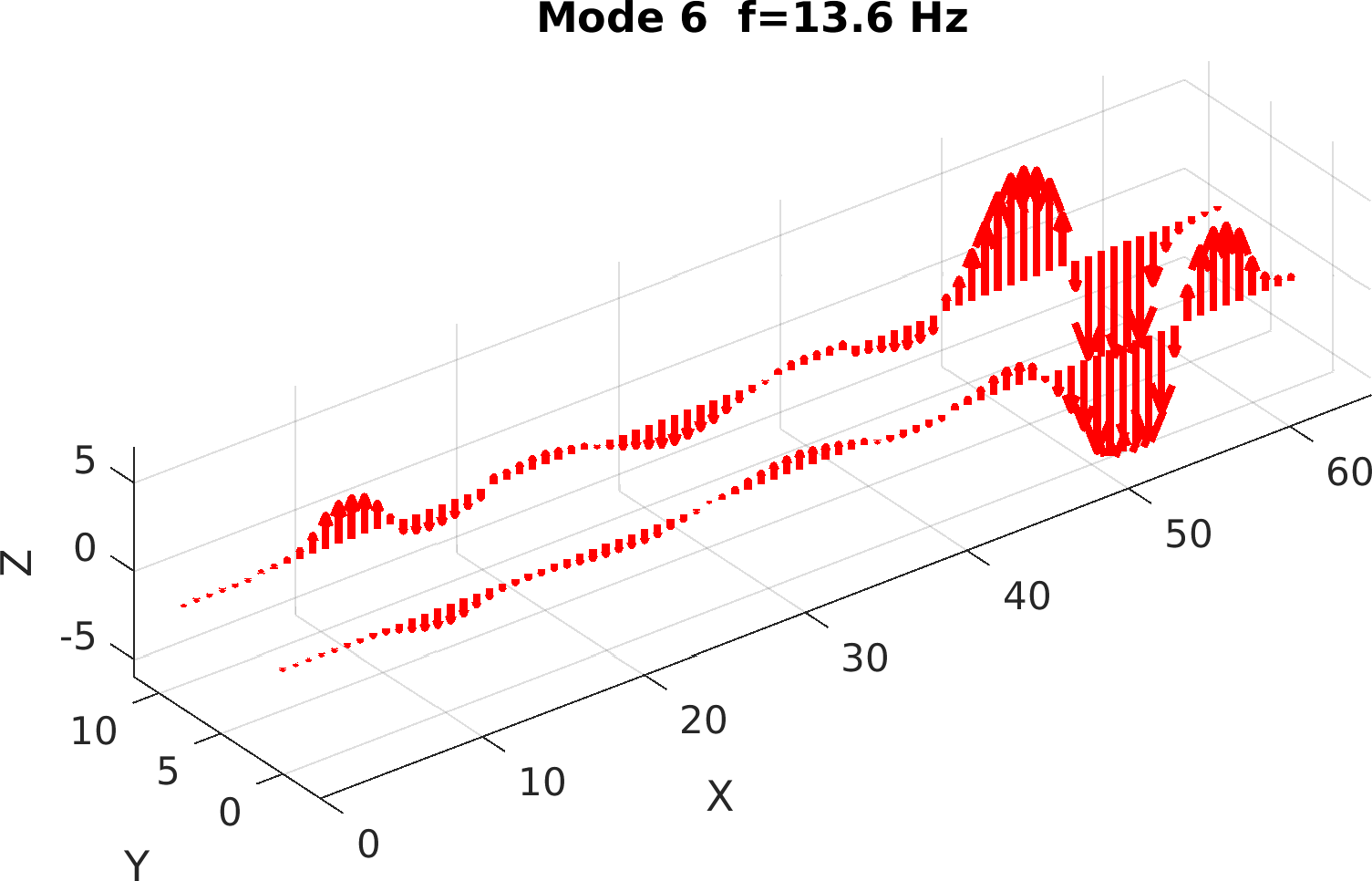}\hfill
\includegraphics[width = 0.33\textwidth]{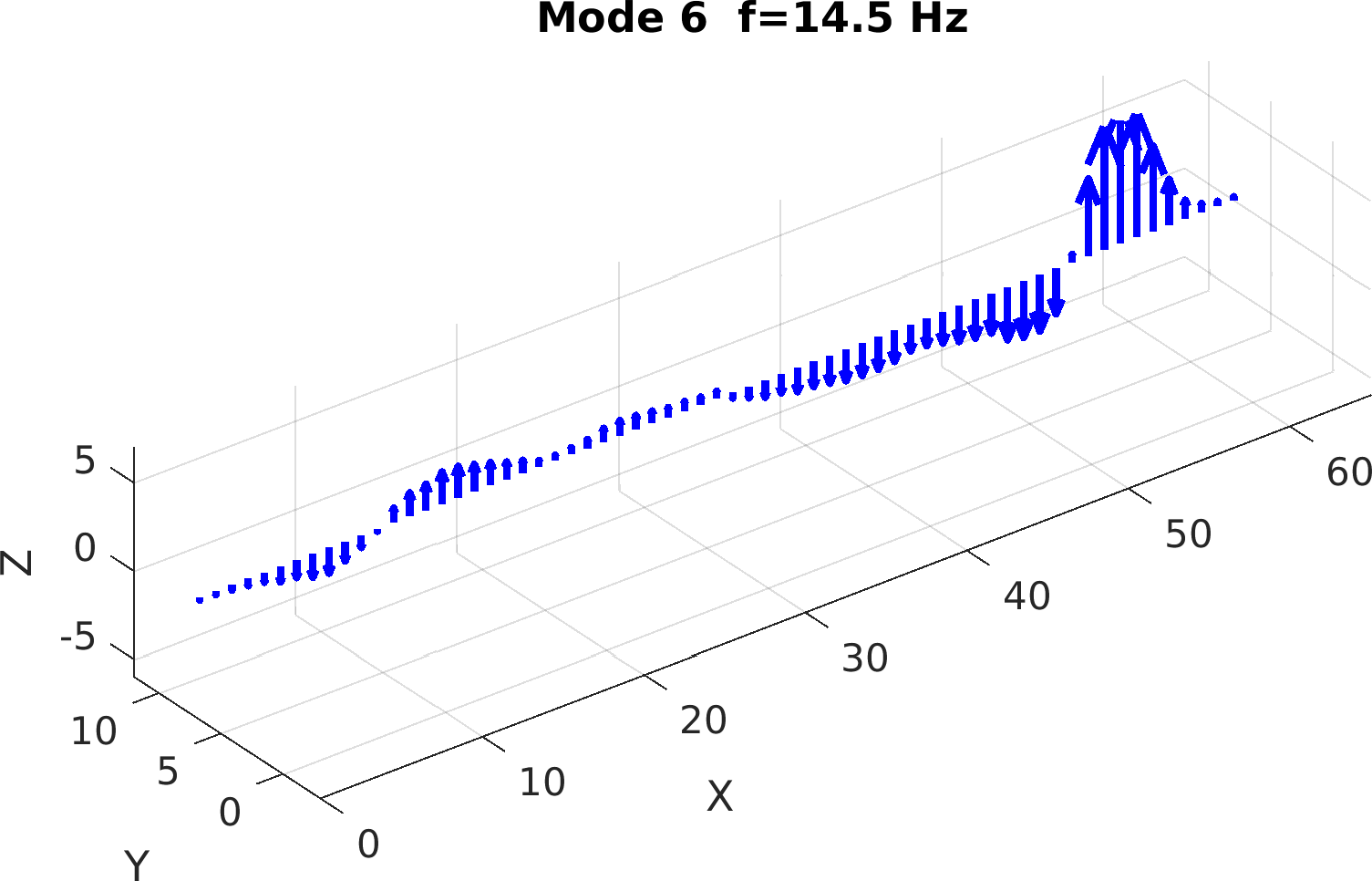}
\includegraphics[width = 0.33\textwidth]{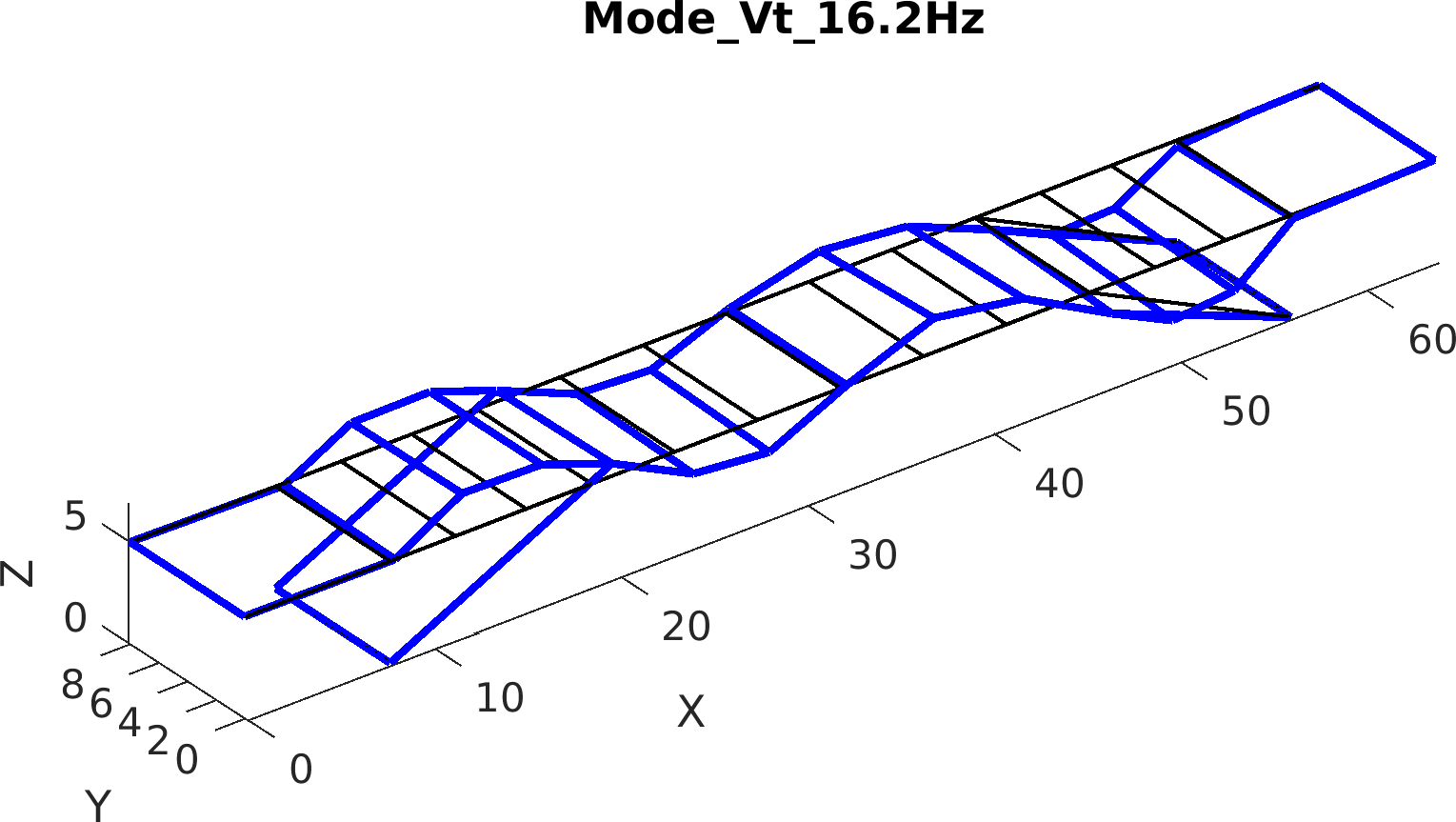}\hfill
\includegraphics[width = 0.33\textwidth]{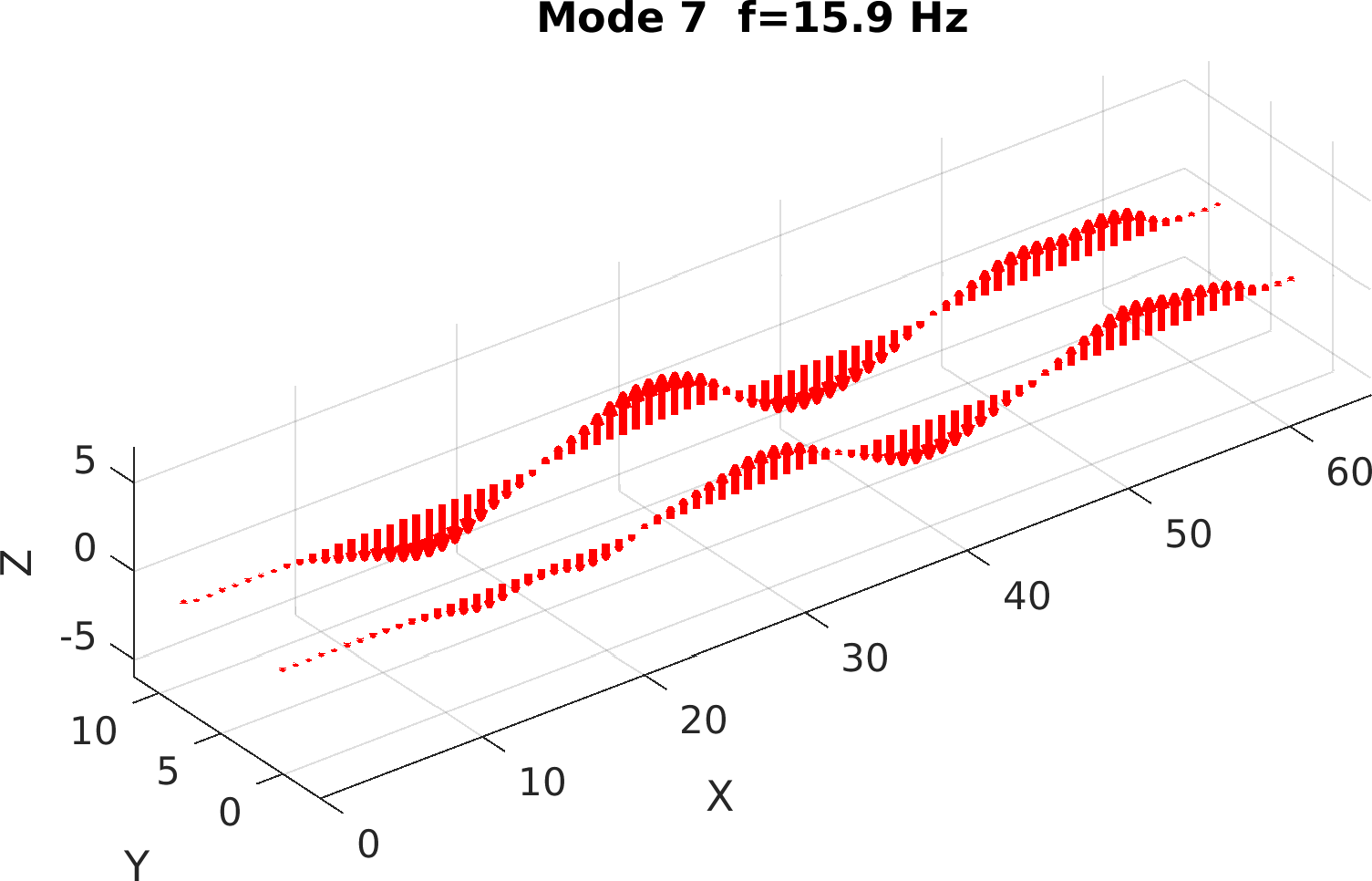}\hfill
\includegraphics[width = 0.33\textwidth]{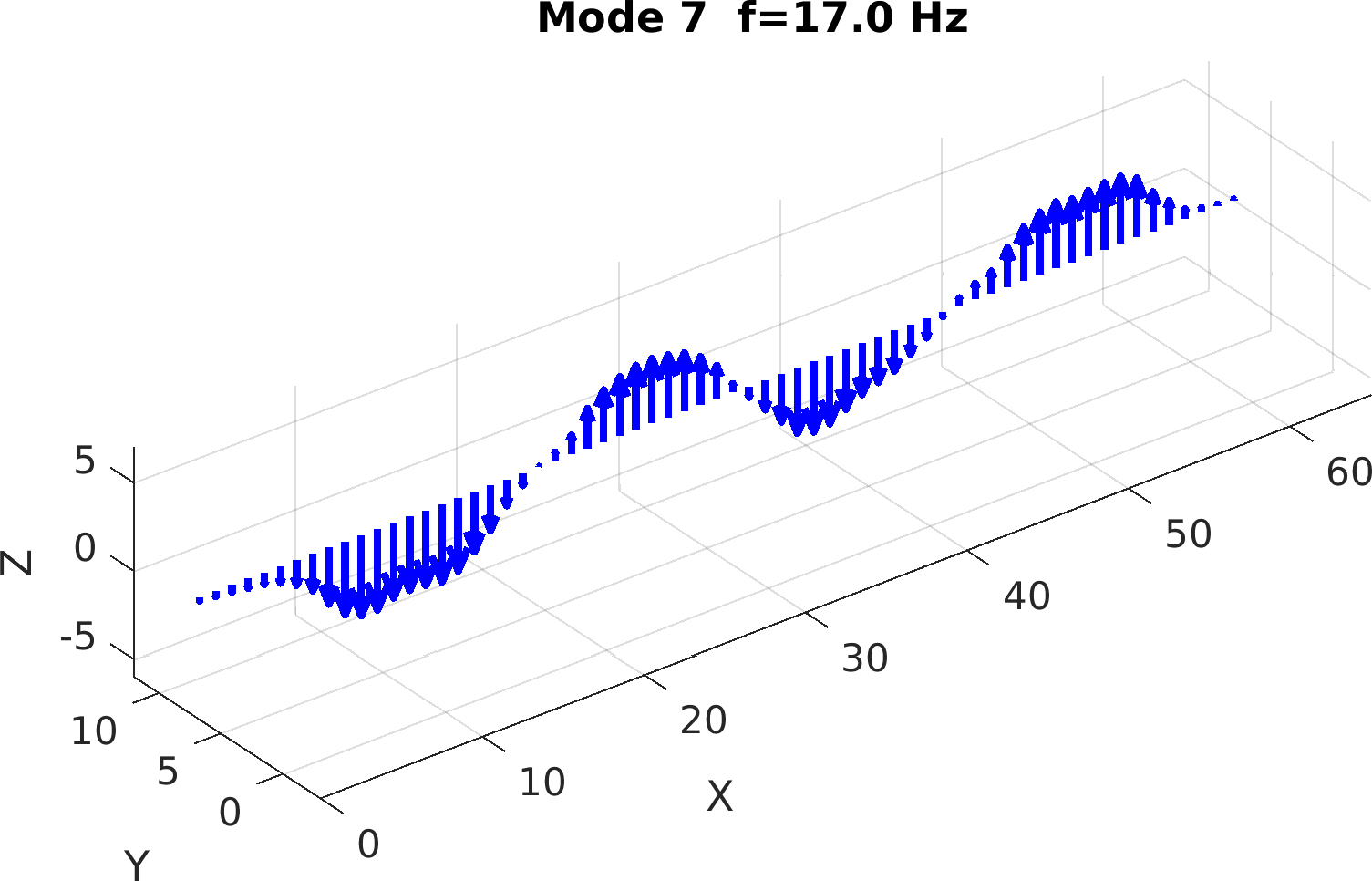}
\includegraphics[width = 0.33\textwidth]{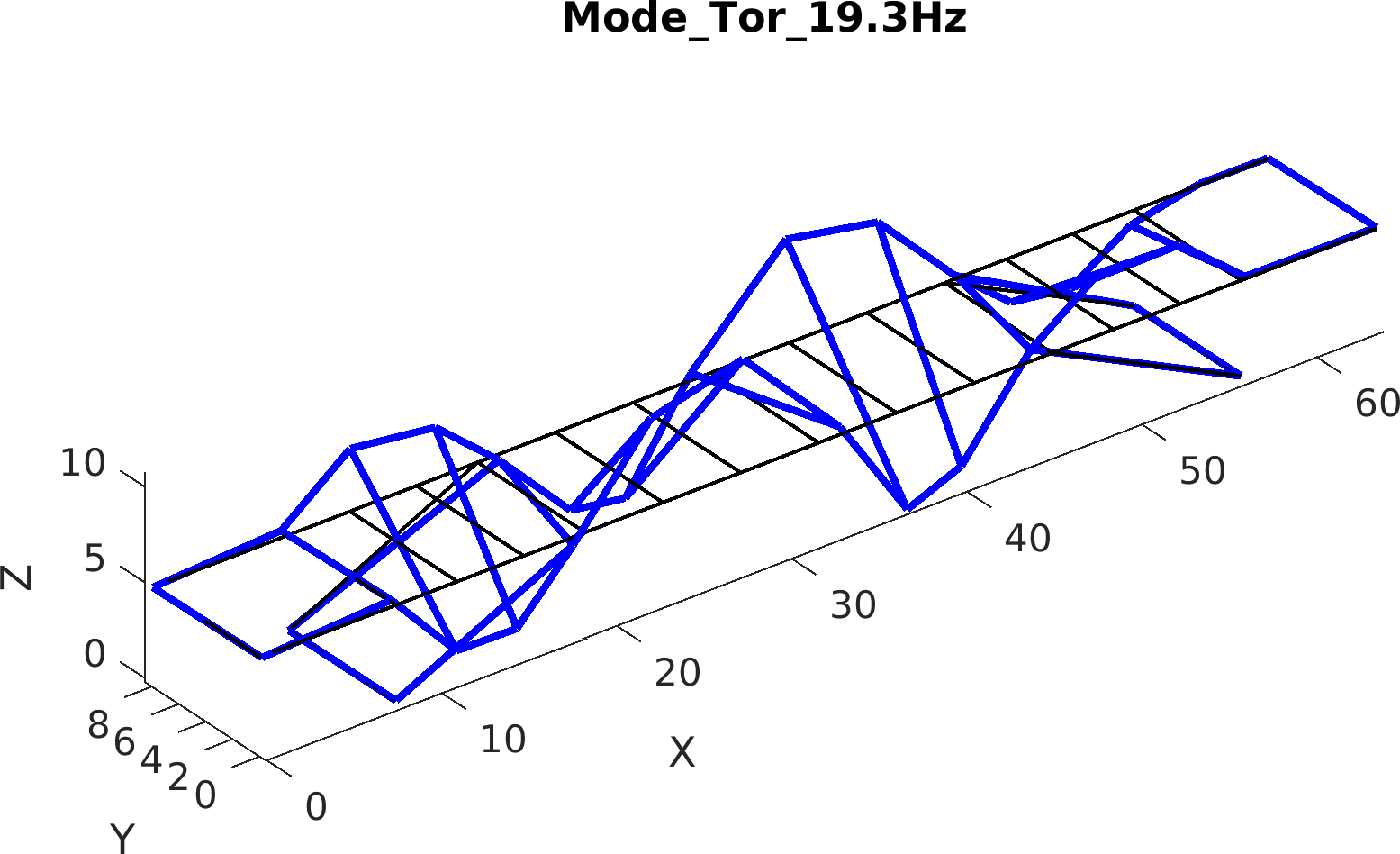}
\includegraphics[width = 0.33\textwidth]{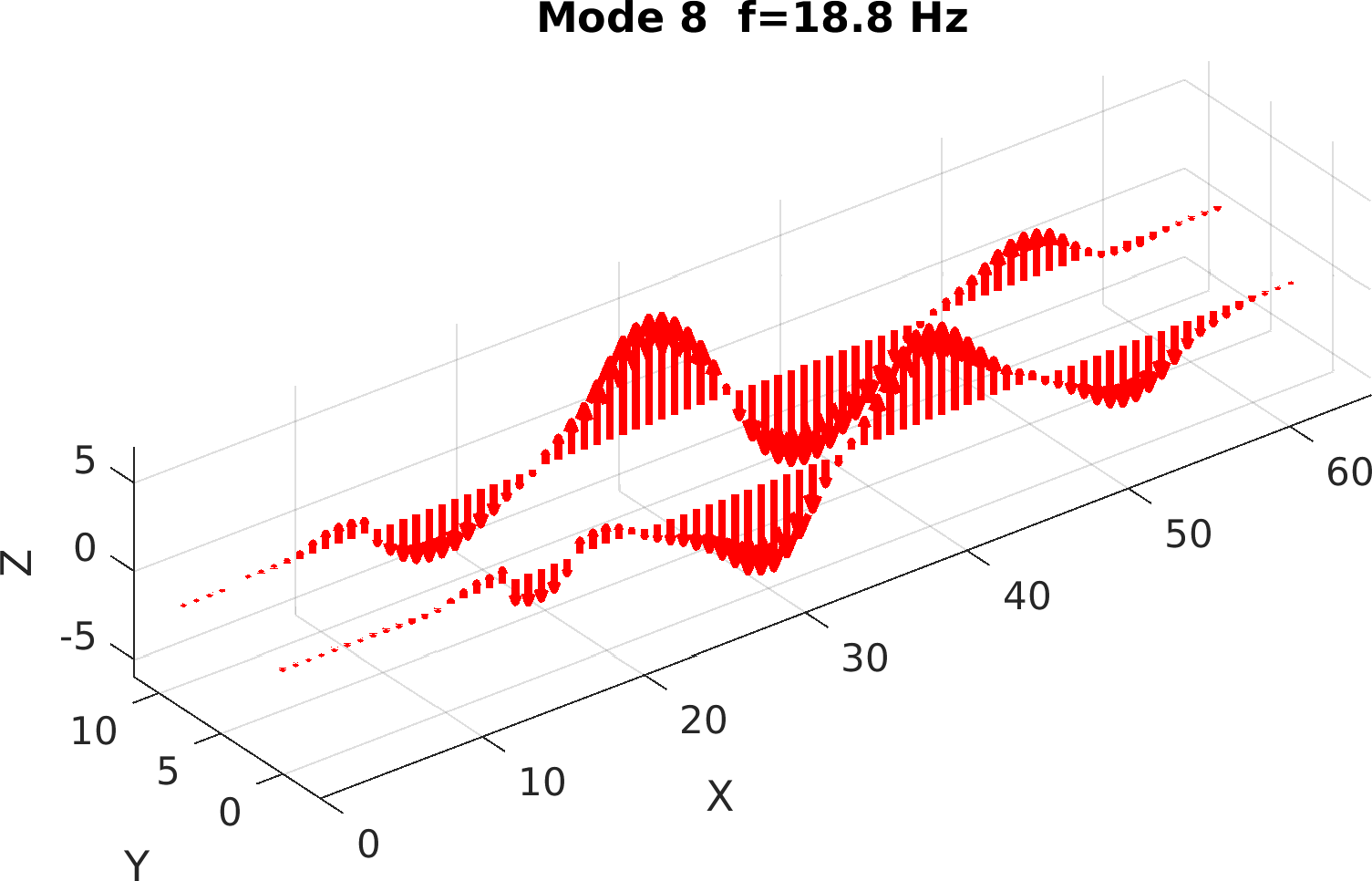}\hfill
\includegraphics[width = 0.33\textwidth]{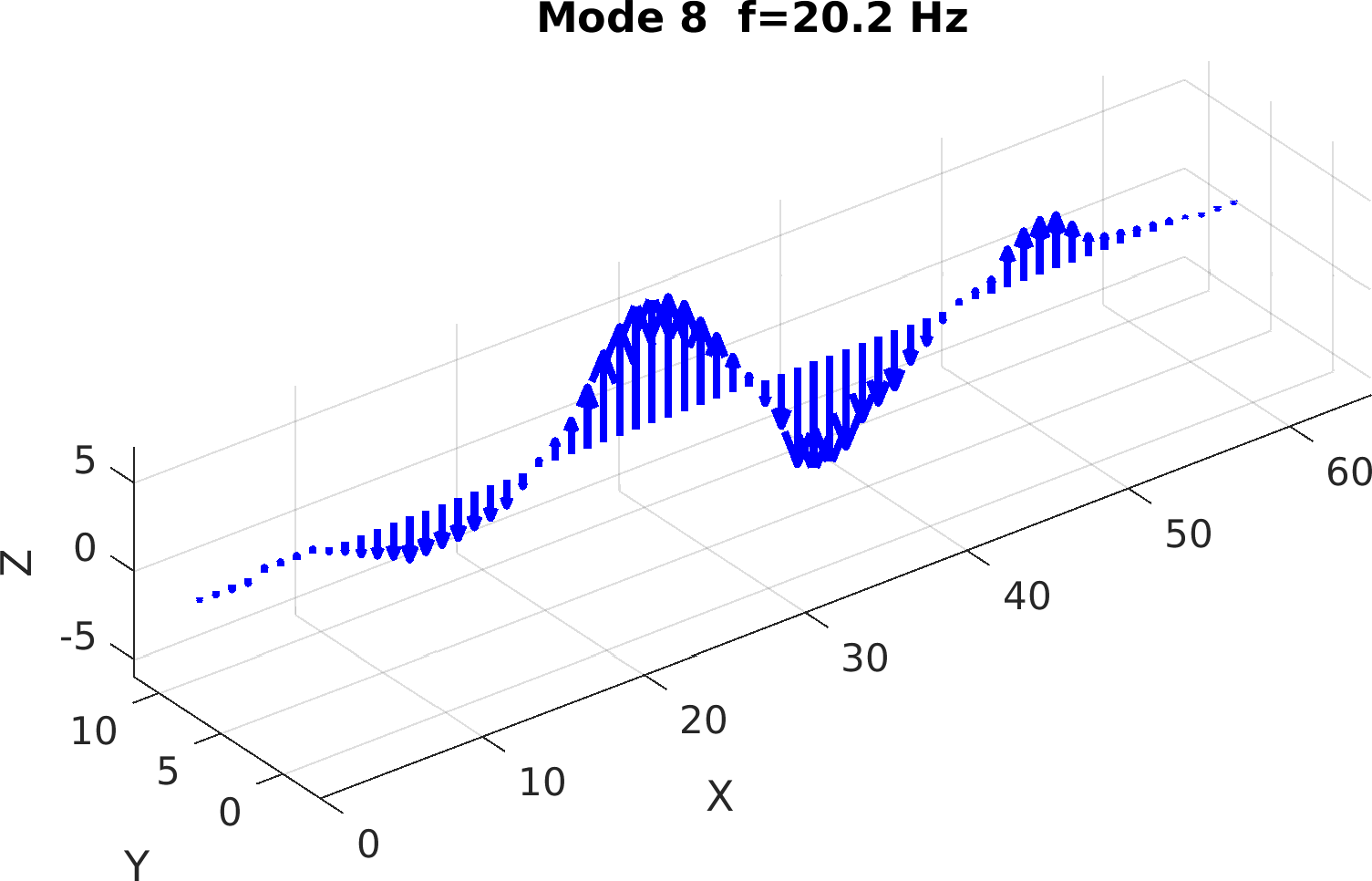}
\caption{Comparison of modal shapes of the strut-frame bridge as inferred from the AVT seismometer data (top row) and DAS data (bottom row). a) torsional mode at 13.6 Hz, b) 2nd order vertical (symmetrical) mode at 16 Hz, and c) 2nd order torsional (anti-symmetrical) mode around 19 Hz.}    
\label{fig:comp_modes_DAS}
\end{figure}

For subsequent modal analysis, we selected several spectral peaks, starting from 4.9 Hz to 20 Hz. We have avoided the channels that are not going to be used for the modal shapes, i.e. the ones crossing the street on the bridge deck, where the SNR is not as good as the other channels. The first four modal shapes resulting from this analysis are shown in Fig~\ref{fig:modes_MP}. More modal shapes, corresponding to higher-order modes, are shown in Fig~\ref{fig:comp_modes_DAS}. For visualization, we plot the eigenvectors of the CPSD matrix calculated from the strain-rate signals recorded by the fiber and choose one direction with respect to the main polarization of each mode. We recall that the DAS measurement gives only the longitudinal strain-rate value per sensing point. For instance, for the first frequency at 4.9 Hz, we use the transverse direction as we already know that the mode at 4.9 Hz is the first transverse mode from the AVT of 2019. For the following three modes at 7.6 Hz, 8.8 Hz and 12.6 Hz we use just the vertical direction as they correspond to vertical bending and torsional modes of the deck.  

One of the striking features that we recognize while interpreting the modal shapes for DAS survey of summer 2022, is the absence of any particular longitudinal motion, even though the optic fiber was laid out exactly following the longitudinal direction of the deck. In fact, at first glance, it may seem as if the first frequency peak at 4.9 Hz is related to this longitudinal mode, but as it will become clear later, this first longitudinal mode is simply missing in the 2022 survey; this is likely due to the longitudinal motion of the bridge being impeded during summer, when the ambient temperatures are higher. Moreover, the transversal mode that is found at 4 Hz in winter time, being found at 4.9 Hz during the summer 2022 survey. By comparing the different columns of Fig~\ref{fig:modes_MP}, we identify the first mode as the transverse bending of the deck, while the second, third, and fourth modes present important vertical motion, and are therefore related to vertical bending modes. What is clearly missing is the longitudinal mode that was identified at 5.1 Hz in winter 2019 during the classical seismometer-based OMA survey (second row in Fig~\ref{fig:modes_MP}). Another fact that supports this interpretation relies on the analysis of the late DAS survey of winter 2024, when the transverse mode is again found at 4.1 Hz, and the longitudinal mode (with the important vertical buckling both in the central part and at the abutments) is found at 5.2 Hz (see second row in Fig~\ref{fig:modes_MP}).


Regarding higher-order modes, the comparison between modes identified during winter 2019 (seismometer-based AVT), summer 2022 and winter 2024 (DAS data) is shown in Fig~\ref{fig:comp_modes_DAS}. Again, for these four modes, we decided to use the vertical direction as the main direction of motion to interpret the longitudinal strains due to the main vertical bending of the deck. It must be stressed that this could be done as contiguous measurements (and subsequent analysis) with seismometers had already been done on the bridge. The association of longitudinal strain-rate to exclusively vertical motion of the deck is a strong but acceptable hypothesis. In any case, from an SHM perspective, it is important to mention that the AVT OMA identification always requires a kind of interpolation of the modal shapes between sensors spaced tens of meters apart. On the other hand, DAS allows us to obtain a submetric scale (0.8 m in our case) measurement of the strain-rate, without the need of any interpolation of the estimated structural motion. 

\section*{Discussion: seasonal effect on boundary conditions}

\begin{figure}[t]
a)\includegraphics[width = 0.3\textwidth]{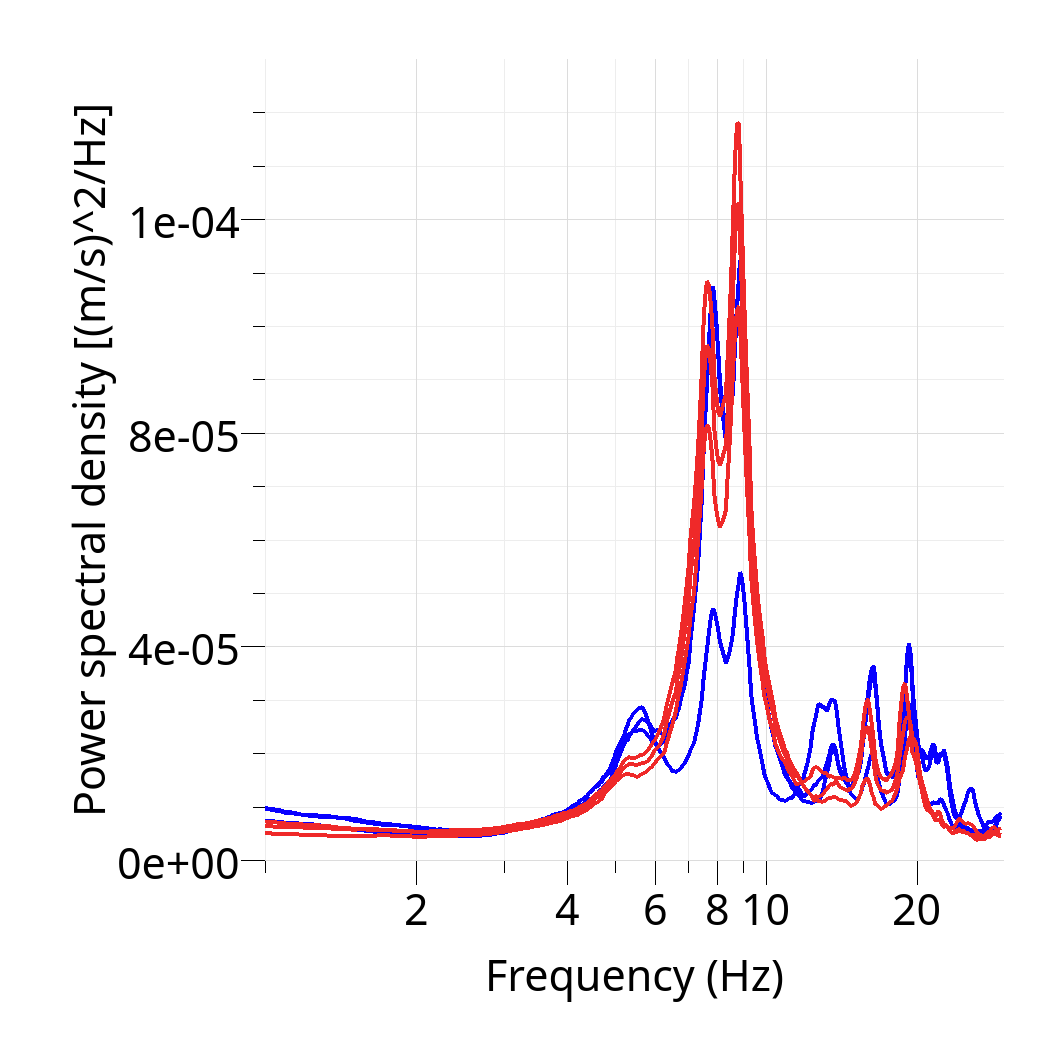}\hfill
b)\includegraphics[width = 0.3\textwidth]{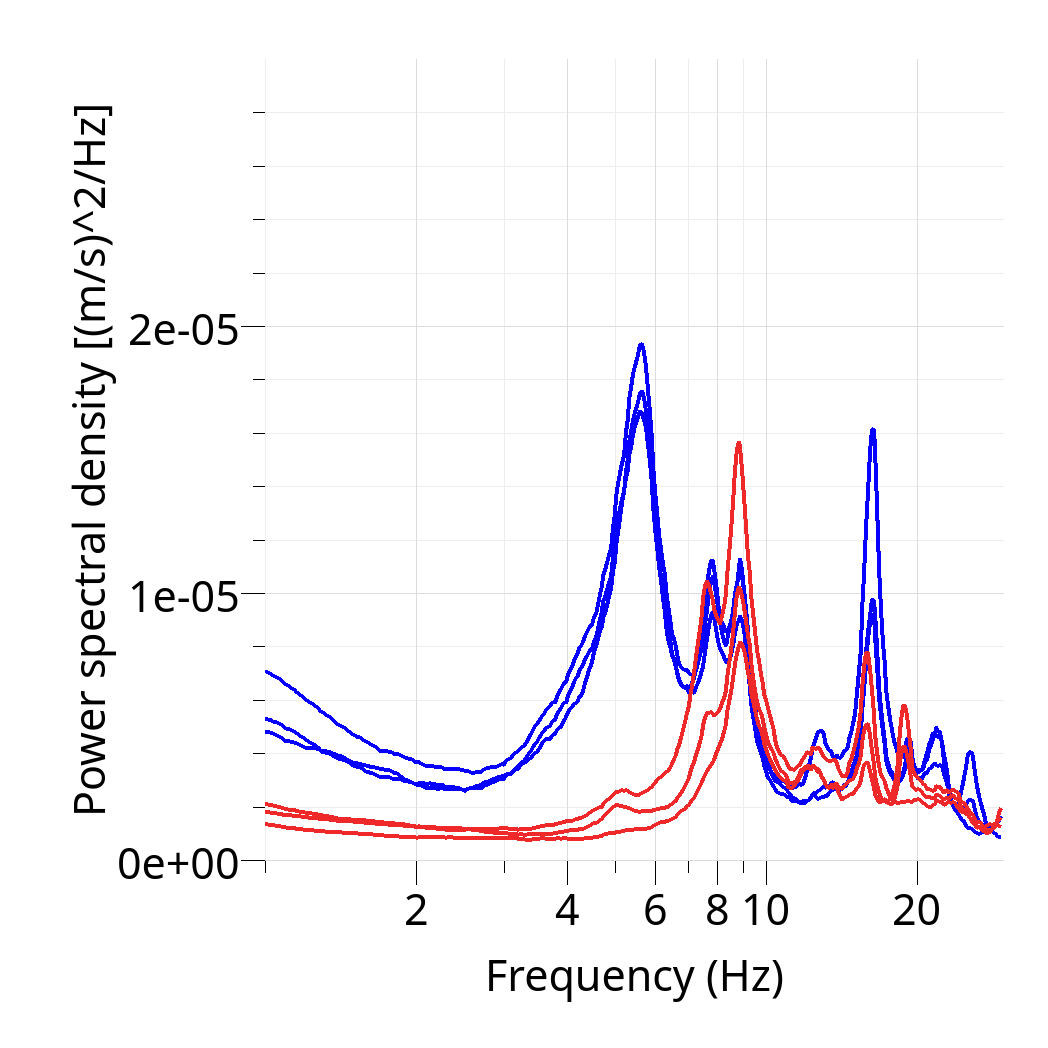}\hfill
c)\includegraphics[width = 0.3\textwidth]{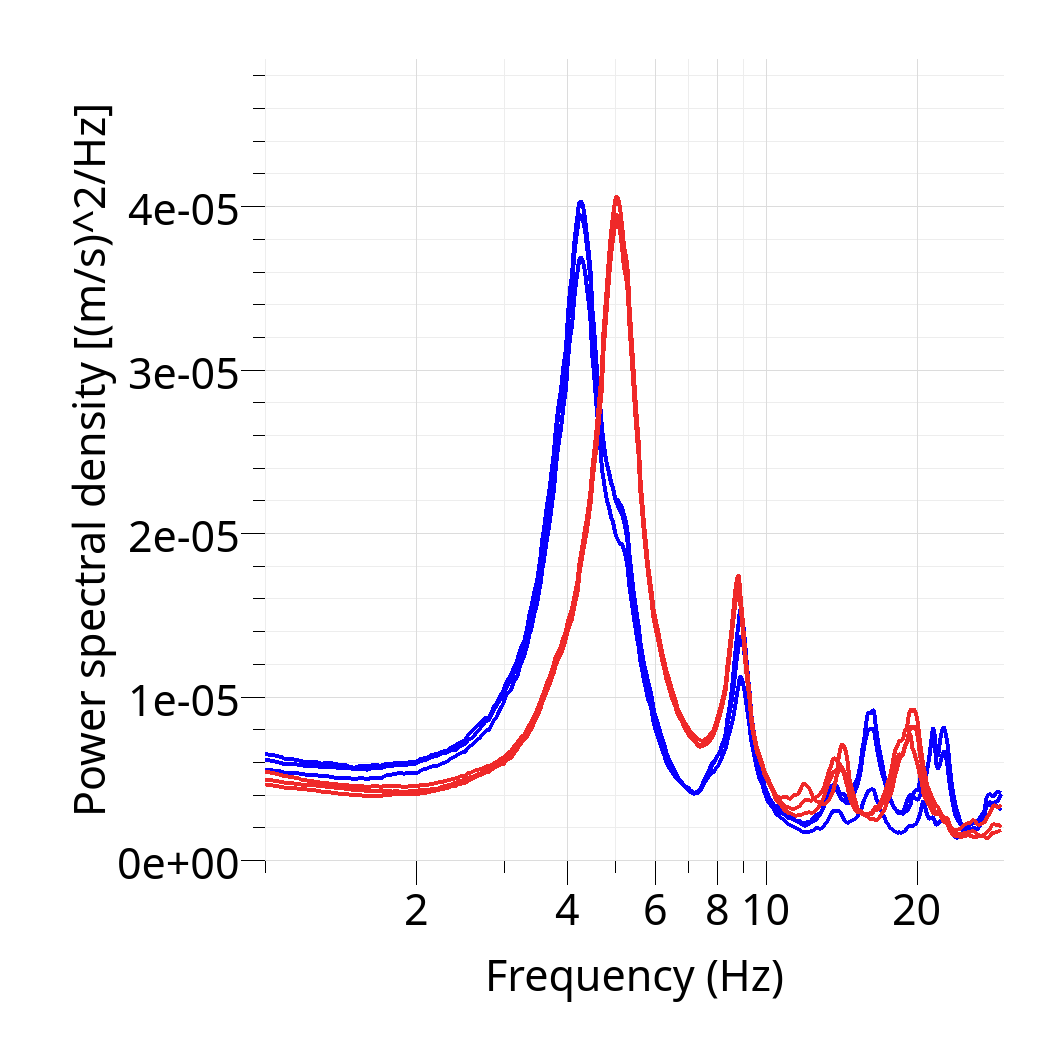}
\caption{A comparison of seismometer spectra between the winter 2019 (blue) and summer 2022 (red) surveys. Each curve corresponds to the mean value of one hour of data. a) Vertical component of velocity b) longitudinal component c) transversal component.}    
\label{fig:spectra_ete_hiver}
\end{figure}

What is apparent from the results reported above is that there are marked differences in the structural dynamics of the strut-frame bridge between winter (January 2019 and 2024) and summer (June 2022). These differences are also seen when only the seismometer data recorded during these two surveys are compared, indicating that this is not just an artifact or limitation of the DAS measurements. A likely physical origin of the observed differences is the seasonal change in temperature (more than 20~$^{\circ}$C between both surveys). This temperature difference may affect the resonance frequencies of the whole soil-structure system, particularly modifying the boundary conditions \citep{mertlichSTRUCTURALHEALTHMONITORING2008,salehiIdentificationBoundaryConditions2022}. Looking at the power spectral densities from the seismometer data recorded at the central span of the bridge in detail (Fig.~\ref{fig:spectra_ete_hiver}), we see that some of the peaks that are observed in winter (almost) disappear during the summer; this is especially apparent for the first transversal and longitudinal modes. The transversal mode at 4 Hz in winter is shifted to approximately 5 Hz in summer (20\% difference, Fig~\ref{fig:spectra_ete_hiver}c), while the first longitudinal mode at 5.1 Hz, clearly visible during the winter survey in 2019, is almost absent during the summer survey in 2022 (tiny peak in Figure \ref{fig:spectra_ete_hiver}b, red curve). These two behaviors can be explained by the thermal expansion of the deck in summer, possibly causing the thermal joints to block at both bridge abutments, drastically decreasing the longitudinal motion. At the same time, the bridge becomes stiffer in the transverse direction, as suggested by the frequency of the transverse mode increasing from 4 Hz to around 5 Hz in summer. Not surprisingly, the vertical modes at 7.6 Hz and 8.7 Hz are less affected by this seasonal change of boundary conditions, and the spectral peaks present similar amplitudes in both winter and summer periods. It is interesting to note that in the case of high-rise RC buildings  \citep{clintonObservedWanderNatural2006a,yuenAmbientInterferenceLongterm2010}, frequencies and temperature have already been shown to be positively correlated, mainly due to the competing behavior of thermal expansion (closing of cracks) and the reduction of the Young modulus while the temperature increases. On the other side, the comparison of the spectral peaks of high-order modes ($>$ 8 Hz) of both DAS campaigns in summer 2022 and winter 2024, shows a slight stiffening of the structural behavior in the latter, as expected due to thermal contraction.     

In summary, seasonal temperature variations modify not only the resonance frequencies but also the modal behavior, even causing the disappearance of one specific (longitudinal) mode caused by thermal expansion of the deck. Environmental effects have previously been reported in several other studies of bridge structures: in the case of cable-stayed bridges \citep{niCorrelatingModalProperties2005,cheynetTemperatureEffectsModal2017}, curved post-tensioned concrete bridges  \citep{liuEffectTemperatureModal2007,salvermoserStructuralMonitoringHighway2015} and also at the laboratory scale  \citep{xiaLongTermVibration2006}. Long-term monitoring databases have recently been published with several other examples  \citep{xiaTemperatureEffectVibration2012,zhouSummaryReviewCorrelations2014,daroThermalEffectsBridges2023}. However, all of these focused on the environmental effects (mostly thermal expansion) on the dynamic properties of the bridge decks and cables (for suspension bridges), while almost none considered the effect on modifying the boundary conditions at the abutments. Recent work, as well as the present study, clearly shows that frequency variations of over 20\% can be expected for different boundary conditions in long-span suspension bridges \citep{liEffectBoundaryConditions2010} and curved I-girder bridges \citep{mertlichDynamicStaticBehavior2007,mertlichSTRUCTURALHEALTHMONITORING2008}, particularly for the lowest (transverse and longitudinal) modes. The recent work of \cite{weiThermalEffectDynamic2024} based on laboratory-scale experimental results clearly states that the variation in bearing stiffness caused by thermal effects can significantly influence the dynamic characteristics of bridges. This must be taken into account in any future SHM survey to be implemented on the bridge.

Based on this interpretation of environmental factors that affect the strut-frame bridge under study, long-term continuous monitoring of the structure would be a pertinent objective for future investigations. The agreement between DAS and seismometer data obtained during the 2022 survey warrants the use of DAS for this purpose; rather than arrange a permanent deployment of seismometers or strain gauges, which is both costly and highly impractical, an optical fiber cable (having a nominal cost of the order of 1~€/m) could be permanently fitted to the bridge deck. Optical fibers for telecommunications usually go through bridges, which might be an easier alternative. This would allow for both continuous and intermittent surveys of the bridge structural dynamics and a more precise observation of seasonal transients in the resonance induced by temperature changes. 

\section*{Conclusions}

We present here one of the first reported cases using Distributed Acoustic Sensing (DAS) for the dynamic characterization of bridges, in this case an overpass highway bridge located in southeastern France. By deploying a fiber optic cable along both sides of the bridge deck, we were able to record the stationary resonance frequencies of the structure with DAS with extremely high spatial resolution (less than 1~m). By performing the Frequency Domain Decomposition of the strain-rate recordings, we extracted the modal shapes of the first eight eigenmodes of the structure. Compared to the results obtained with conventional seismometers in a previous OMA survey in winter 2019, we find marked seasonal effects induced by a difference in temperature of $\sim$20 $^{\circ}$C. We observe a 20\% change in the frequency of the first transverse mode of the deck (from 4 Hz to 4.9 Hz in winter and summer periods, respectively) and the disappearance of the first longitudinal mode at 5.1 Hz in the summer period. The disappearance of this mode is potentially related to a change in the boundary conditions, with thermal expansion of the deck causing the joints to lock up at both bridge abutments. This result has been confirmed after the second DAS survey of winter 2024, when the longitudinal mode has been again identified at 5.2 Hz. On the other hand, high-order modes (mainly vertical bending and torsion of the deck) show the classical behavior of thermal contraction in cold winter weather, which causes the slight stiffening of the structure. These observations have natural implications for long-term structural health monitoring of bridge structures using DAS and the design of future structures. Lastly, the low cost and ease of deployment of the fiber-optic cable (the cable can be left in place between campaigns), as well as the reliability and high spatial resolution of the DAS measurements, warrant the use of this technology in future structural health monitoring surveys, including long-term studies on the seasonal effects of structural dynamics.








\subsection*{Compliance with Ethical Standards:}

Funding: this work was financially supported by the ANR-19-CE04-0011-01 MONIDAS (Natural Hazard Monitoring using Distributed Acoustic Sensing).\\
Conflict of Interest: The authors declare that they have no conflict of interest.\\
Data availability : Data used in the present paper will be available upon reasonable request to the authors.\\

\subsection*{Contributor Roles}
E.D. Mercerat : Conceptualization
Data acquisition and curation, Formal analysis, Methodology, Writing – original draft, review and editing\\
M van den Ende : Conceptualization
Data acquisition and curation, Methodology, writing – review and editing.\\
A. Sladen : Conceptualization
Data acquisition and curation, Methodology, writing – review and editing.\\
V Carrillo Barra, Data acquisition, writing – review and editing.\\
J Pel\'aes Qui\~nones, Data acquisition, writing – review and editing.\\
D Mata Flores, Data acquisition, writing – review and editing.\\
Ph. Langlaude, Data acquisition, writing – review and editing.\\
D Nziengui : Funding acquisition, writing review and editing\\
O Coutant : Funding acquisition, Project administration, writing review and editing

\bibliography{mybiblio}

\end{document}